\documentclass[superscriptaddress, twocolumn,prd,preprintnumbers,amsmath,amssymb,nofootinbib]{revtex4}

\usepackage{graphicx}
\usepackage{dcolumn}
\usepackage{bm}
\usepackage{epsfig}
\usepackage{amssymb}
\usepackage{amsmath}

\newcommand{\aap}{Astron. Astrophys.}

\newcommand{\apjl}{Astrophys. J. Lett.}
\newcommand{\apjs}{Astrophys. J. Suppl. Ser.}

\newcommand{\jcap}{JCAP}

\newcommand{\mnras}{Mon. Not. R. Astron. Soc.}

\newcommand{\plb}{Phys. Lett. B}

\newcommand{\lsim}{\mathrel{\hbox{\rlap{\lower.55ex\hbox{$\sim$}} \kern-.3em \raise.4ex \hbox{$<$}}}}
\newcommand{\gsim}{\mathrel{\hbox{\rlap{\lower.55ex\hbox{$\sim$}} \kern-.3em \raise.4ex \hbox{$>$}}}}
\newcommand{\beq}{\begin{equation}}
\newcommand{\eeq}{\end{equation}}
\newcommand{\beqa}{\begin{eqnarray}}
\newcommand{\eeqa}{\end{eqnarray}}
\newcommand{\drm}{\mathrm{d}}
\newcommand{\grad}{\vec{\nabla}}

\newcommand{\gam}{\Gamma_\phi}
\newcommand{\gamo}{\Gamma_\phi}
\newcommand{\gamt}{\tilde{\Gamma}_\phi}
\newcommand{\rhos}{\rho_\phi}
\newcommand{\rhor}{\rho_r}

\newcommand{\rhom}{\rho_\mathrm{dm}}
\newcommand{\rhost}{\tilde{\rho}_\phi}
\newcommand{\rhort}{\tilde{\rho}_r}
\newcommand{\rhomt}{\tilde{\rho}_\mathrm{dm}}
\newcommand{\dels}{\delta_\phi}
\newcommand{\delr}{\delta_r}
\newcommand{\delm}{\delta_\mathrm{dm}}
\newcommand{\thes}{\theta_\phi}
\newcommand{\ther}{\theta_r}
\newcommand{\them}{\theta_\mathrm{dm}}
\newcommand{\thest}{\tilde{\theta}_\phi}
\newcommand{\thert}{\tilde{\theta}_r}
\newcommand{\themt}{\tilde{\theta}_\mathrm{dm}}
\newcommand{\scrih}{\frac{\dot{a}}{a}}
\newcommand{\tilk}{\tilde{k}}
\newcommand{\kdec}{k_\mathrm{RH}}
\newcommand{\kdect}{\tilde{k}_\mathrm{RH}}
\newcommand{\adec}{a_\mathrm{RH}}
\newcommand{\keq}{k_\mathrm{eq}}
\newcommand{\vdec}{\langle v_\mathrm{RH} \rangle}
\newcommand{\mdec}{M_\mathrm{RH}}
\newcommand{\lfsh}{\lambda_\mathrm{fsh}}
\newcommand{\kfsh}{k_\mathrm{fsh}}
\newcommand{\mfsh}{M_\mathrm{fsh}}
\newcommand{\delcoll}{\delta_c}

\begin{document}

\title{Reheating Effects in the Matter Power Spectrum and Implications for Substructure}

\author{Adrienne L. Erickcek}\email{erickcek@cita.utoronto.ca}
\affiliation{Canadian Institute for Theoretical Astrophysics, University of Toronto, 60 St. George Street, Toronto, Ontario M5S 3H8, Canada}
\affiliation{Perimeter Institute for Theoretical Physics, 31 Caroline St. N, Waterloo, Ontario N2L 2Y5, Canada}

\author{Kris Sigurdson}\email{krs@physics.ubc.ca}
\affiliation{Department of Physics and Astronomy, University of British Columbia, Vancouver, BC V6T 1Z1, Canada}

\date{\today}

\begin{abstract}
The thermal and expansion history of the Universe before big bang nucleosynthesis is unknown.
We investigate the evolution of cosmological perturbations through the transition from an early matter
era to radiation domination.
We treat reheating as the perturbative decay of an oscillating scalar field into relativistic plasma and cold dark matter.
After reheating, we find that subhorizon perturbations in the decay-produced dark matter density are significantly enhanced, while subhorizon radiation perturbations are 
instead suppressed. 
If dark matter originates in the radiation bath after reheating, this suppression may be the primary cutoff in the
matter power spectrum.  Conversely, for dark matter produced nonthermally from scalar decay, enhanced perturbations can drive structure formation
during the cosmic dark ages and dramatically increase the abundance of compact substructures. For low reheat temperatures, we find that as much as 50\% 
of all dark matter is in microhalos with $M \gtrsim 0.1M_\oplus$ at $z \simeq 100$, compared to a fraction of $\sim\!10^{-10}$ in the standard case. 
In this scenario, ultradense substructures may constitute a large fraction of dark matter in galaxies today.

\end{abstract}
   
\maketitle

\section{Introduction}

The standard cosmological model accounts for a remarkable span in the Universe's history, from the epoch of big bang nucleosynthesis (BBN)
to the era of structure formation to our present dark-energy driven abyss.  In the standard paradigm, inflation \cite{Guth80, AS82, Linde82} cures the horizon and flatness problems of the hot big bang and generates the spectrum of curvature perturbations needed for the formation of large-scale structure.  Inflation ends when the kinetic energy of the inflaton field is no longer small compared to its potential energy.  At some point after the end of inflation, the inflaton decays, and the Universe becomes radiation dominated.  

The inflationary paradigm does not uniquely specify the reheating and thermal history of the Universe, however.   The only constraint on the Universe's temperature at the beginning of the radiation-dominated era comes from the thermal production of neutrinos; lowering this temperature lowers the neutrino abundance \cite{KKS99, KKS00, Han04, IKT05}, which has ramifications for BBN, the cosmic microwave background (CMB), and large-scale structure.  If the Universe is radiation dominated at a temperature of $\sim\!3\,{\rm MeV}$, then thermal neutrino production is sufficient to produce the observed abundances of light elements \cite{KKS99, KKS00, Han04, IKT05} and the observed CMB and matter power spectra \cite{IKT07,dBPM08}, provided that the Universe is baryon-asymmetric at this temperature with a baryon-to-photon ratio of $\eta \simeq 6 \times10^{-10}$ \cite{IKT05, wmap7param}.  While challenging, it is possible to generate this baryon asymmetry if the Universe is not radiation dominated until it reaches a temperature of $\sim\!\!\!10\,{\rm MeV}$ (e.g.,~\cite{ADS10,DMST2010}).

At the end of inflation it is usually assumed that the inflaton can be described as a coherently oscillating field that eventually decays and transfers its energy to a radiative plasma \cite{ASTW82, Turner83, TB90, KLS94, KLS97, DFKPP06, ABCM10}.  The inflaton is the first dominant energy component of the Universe and its decay could be the start of the standard hot big bang --- or this could be a transient epoch.  It is possible that a spectator field during inflation comes to dominate the energy density of the Universe after the inflaton decays; this scenario is a generic consequence of stabilized moduli in string theories \cite{CCQR93, BKN94, AKK10}, and it is how the initial curvature perturbations are created in the curvaton model \cite{Mollerach90, LM97, LW02, MT01}.
In this case, the second scalar field's decay supersedes the inflationary reheating epoch, and its fluctuations set the spectrum of adiabatic perturbations.  Clearly, this process could occur multiple times, with the caveat that thermal relics may survive subsequent reheatings.  

The final round of reheating takes particular importance because it generates the thermal plasma of the Universe and fixes the spectrum of perturbations on scales greater than the horizon size during the last reheating epoch.   Physical processes that occur at temperatures larger than the temperature of the radiation bath following the last reheating epoch (the reheat temperature) will be affected by the altered expansion history of the Universe prior to reheating.  The impact of a low reheat temperature on the relic abundance of WIMP 
dark matter has been studied extensively \cite{KT90, CKR99, GKR01, FRS03, Pallis04, GG06, DIK06, GGSY06};  a low reheat temperature reduces the relic abundance of thermal dark matter, but the abundance may be enhanced if the dark matter is also produced nonthermally.  A low reheat temperature also relaxes constraints on axion \cite{GKR01, GSK08, VG10} and neutrino \cite{GKR01, GKRST01,Yaguna07} dark matter.   This prior work on low-temperature reheating has only considered its impact on the homogeneous content of the Universe, but deviations from radiation-dominated expansion also affect density perturbations.  
In this paper we consider how the evolution of the Universe prior to the creation of the final radiation bath influences perturbations on scales that enter the horizon before the last reheating epoch.

A homogeneous scalar field that rapidly oscillates around the minimum of a quadratic potential has the same dynamics as a pressureless fluid of non-relativistic matter --- its energy density scales as $a^{-3}$ \cite{Turner83}.  Moreover, it has been shown that the correspondence between a rapidly oscillating scalar field and a matter fluid extends to the perturbations \cite{KMZ85, JK08, JLM10a, EFG10}.  In particular, subhorizon perturbations in the scalar field's energy density with $k_\mathrm{phys} < \sqrt{3Hm_\phi}$, where $m_\phi$ is the mass of the scalar field, grow linearly with the scale factor while the scalar field dominates the Universe \cite{JLM10a, EFG10}.   If the scalar oscillates for a sufficiently long time before decaying, density perturbations in the scalar field become nonlinear and could produce gravitational waves \cite{AW09, JLM10b}.  Longer wavelength perturbations enter the horizon later and remain linear prior to reheating.  Nevertheless, the growth of these perturbations during an early scalar-dominated phase may have a significant impact on the subsequent growth of structure.

To assess effects of an early ``matter"-dominated epoch on the present-day matter power spectrum, we study the evolution of perturbations through the reheating era.   Our analysis applies to any scenario in which the energy that dominates the Universe prior to reheating behaves as a pressureless fluid and decays perturbatively.   This fluid could be composed of metastable non-relativisitic particles, or it could be an oscillating scalar field.   We note that a coherently oscillating scalar field may decay non-perturbatively; during preheating, the oscillations of the scalar field lead to resonant particle production that cannot be captured by standard perturbation theory \cite{TB90, KLS94,KLS97}.  A coherently oscillating scalar field still decays perturbatively, however, if its couplings to other fields are sufficiently small that preheating is unimportant.  Specifically, our analysis applies to the decay of coherent scalar fields in the narrow resonance regime; due to the Universe's expansion, preheating is inefficient in this regime, and the scalar's decay can be treated perturbatively \cite{KLS97}.

Inspired by models in which dark matter is a decay product of a string modulus \cite{MR00, AKWK09}, the inflaton \cite{AD02a, AD02b}, or the curvaton \cite{GMW04}, we allow the pressureless fluid in our reheating model to decay into cold dark matter particles and radiation.  We focus on modes that are linear during scalar domination; these are the largest scales that enter the horizon prior to radiation domination and thus have the highest potential for observational impact.   We find that the radiation perturbation retains no memory of the scalar perturbation's growth.  On the contrary, the radiation density perturbation on scales that enter the horizon during the ``matter"-dominated phase are dramatically suppressed compared to perturbations that enter the horizon after reheating.  Consequently, dark matter that is coupled to or produced from the radiation after reheating will have a cutoff in the matter power spectrum inherited from the last reheating epoch.   Alternatively, if nearly-decoupled dark matter is produced directly at reheating, the dark matter inherits the enhanced density perturbation that grew prior to reheating.  In this case, we get an enhancement of the small-scale matter power spectrum and an epoch of early small-scale structure formation during the cosmic dark ages.

Our paper is organized as follows:  In Sec.~\ref{sec:threefluid} we discuss our three-fluid model for reheating and its background evolution, and 
we present the equations that govern the evolution of the fluid perturbations and their initial conditions.
In Sec.~\ref{sec:radpert} we describe the evolution of the perturbations in the radiation fluid through reheating.  In Sec.~\ref{sec:matpert} we describe the evolution of the cold dark matter decay product, including the effects of free-streaming, and we derive a new transfer function for the matter perturbation.  In Sec.~\ref{sec:structure} we show that the new small-scale matter power spectrum leads to an abundance of low-mass structures at high redshift.  In Sec.\ref{sec:summary} we summarize our results and briefly discuss the prospects for detecting the numerous compact dark matter subhalos predicted by this reheating scenario.  Appendix \ref{sec:perts_derivation} contains the derivation of the perturbation evolution equations, and the initial conditions for the perturbations are derived in Appendix \ref{sec:perts_ic}.  Finally, Appendix \ref{sec:ABfit} provides expressions necessary to evaluate the matter transfer function, including a scale-dependent growth function for small-scale perturbations.

\section{Three-fluid model for reheating}
\label{sec:threefluid}
Our reheating model consists of a pressureless fluid that dominates the energy density of the Universe before it decays into radiation and dark matter particles.   We assume that the radiation is tightly coupled and behaves like a perfect fluid with $P_{\rm r}=\rhor/3$. 
We also assume that the dark matter particles are nonrelativistic and completely decoupled from the radiation bath from the moment of their creation.   We initially neglect the velocity dispersion of these particles, but in Section \ref{sec:freestream}, we consider the impact of free-streaming by dark matter particles that are created with a nonzero velocity. 

Inspired by the curvaton and inflaton, and to avoid confusion with dark matter produced during reheating, we call the pressureless fluid that initially dominates the Universe a scalar field.  As mentioned in the previous section, an oscillating scalar field behaves like a pressureless fluid, with $\rho \propto a^{-3}$, if the period of the scalar oscillations is much shorter than the Hubble time ($m_\phi \gg H$).   We will assume that the decay of this scalar field is slow enough that the coherence of the scalar oscillations are irrelevant;  in the language of preheating, we are assuming that the scalar decay is the narrow resonance regime in an expanding universe.  If the scalar decay rate $\gam$ satisfies 
${\gam}/{m_\phi} \ll \left({m_\phi}/{m_\mathrm{Pl}}\right)^2$,
then momentum modes of the scalar's decay products redshift through their instability bands too quickly to be significantly excited.\footnote{This condition follows from demanding that  $q \ll H/m_\phi \ll 1$, as in Ref. \cite{KLS97}}   In this case, the decay of the scalar field can be treated perturbatively and the scalar's coherence is irrelevant.   

In this three-fluid model for reheating, the background equations for the energy density $\rhos$ of the scalar field, the radiation density $\rhor$, and the dark matter density $\rhom$ are
\beqa
\frac{\drm}{\drm t} \rhos+3H\rhos &=& -\gam \rhos, \label{bkgdS}\\
\frac{\drm}{\drm t} \rhor+4H\rhor &=& +(1-f)\gam \rhos, \label{bkgdR}\\
\frac{\drm}{\drm t} \rhom+3H\rhom &=& +f\gam \rhos, \label{bkgdM}
\eeqa
where $f$ is the fraction of the scalar's energy that is transferred to dark matter particles.  For a given reheat temperature, there is one value of $f$ that gives the observed dark matter abundance, as discussed below.   In these equations, we have considered only interactions that are linear in the scalar field; such interactions are required for the scalar to decay completely \cite{DFKPP06}.  Nonlinear interactions, which correspond to scalar self-annihilations, would introduce terms proportional to $\rhos^2$ in the equations above.  If one considers an oscillating  scalar field that decays into a massless scalar field $\chi$ via an interaction term proportional to $\phi\chi^2$, the Boltzmann equation averaged over many oscillations implies that $\rho_\chi$ follows Eq.~(\ref{bkgdR}) \cite{BKB10}.

Similar three-fluid models for reheating have been used to calculate the relic abundance of dark matter in low-reheating-temperature cosmologies \cite{CKR99, GKR01, GG06}.  These models also include dark matter self-annihilations and thermal production.  By neglecting such interactions in Eqs. (\ref{bkgdR}) and (\ref{bkgdM}), we are assuming that the dark matter particles produced in scalar decays far outnumber any dark matter particles that may have been produced thermally.  We are also assuming that the velocity-averaged self-annihilation cross section ($\langle \sigma v \rangle$) of the dark matter particles is small enough that the nonthermal dark matter does not self-annihilate.  For any value of $\langle \sigma v\rangle$, these assumptions are valid for sufficiently low reheat temperatures.  If $\langle \sigma v\rangle$ is too large, however, these reheat temperatures may be below 3 MeV and are therefore disallowed by BBN \cite{GG06}. 

With these assumptions, the duration of the radiation-dominated era after reheating is determined by the branching ratio $f$ and does not depend on $\langle \sigma v\rangle$ \cite{GG06}.   If $T_\mathrm{RH}$ is the reheat temperature, and $T_\mathrm{eq}$ is the temperature at matter-radiation equality, then $f\simeq 0.43(T_\mathrm{eq}/T_\mathrm{RH}) (10.75/g_{*S})^{1/3}$, where $g_{*S}$ is the entropy density divided by $(2\pi^2/45)T^3$ and should be evaluated at the reheat temperature.\footnote{We focus on non-relativistic dark matter at reheating. Models with relativistic dark matter at reheating \cite{S2009} need a higher  $f$ for the same late-time abundance, but we do not consider this case.}
From the cosmic microwave background, we know that $T_\mathrm{eq} = 0.75$ eV  \cite{wmap7param}, so \mbox{$f \simeq (6.4 \times 10^{-8}) (\mbox{5 MeV}/T_\mathrm{RH})(10.75/g_{*S})^{1/3}$} is required to produce the observed amount of dark matter.

When numerically solving Eqs. (\ref{bkgdS})-(\ref{bkgdM}) for the evolution of these three fluids and their perturbations, we define an initial time $t_0$ with $a(t_0)\equiv1$ and $H(t_0)\equiv H_1$.   At $t=t_0$, the critical density is $\rho_\mathrm{crit,0} = 3m_\mathrm{Pl}^2H_1^2/(8\pi)$, and we define dimensionless density variables $\rhost \equiv \rhos/\rho_\mathrm{crit,0}$, $\rhomt \equiv \rhom/\rho_\mathrm{crit,0}$, and $\rhort \equiv \rhor/\rho_\mathrm{crit,0}$.  We also define a dimensionless decay rate $\gamt \equiv \gam/H_1$.  We will assume that the scalar field initially dominates the energy density of the Universe.  The scalar field decays when $H \simeq \gam$.  Since the Universe is effectively matter dominated prior to the decay of the scalar field, $H\propto a^{-3/2}$, and the value of the scale factor at the time of scalar decay is $\adec \simeq \gamt^{-2/3}$.  At this time, the Universe transitions from scalar domination to radiation domination.  

\begin{figure}[tb]
 \centering
 \resizebox{3.5in}{!}
 {
      \includegraphics{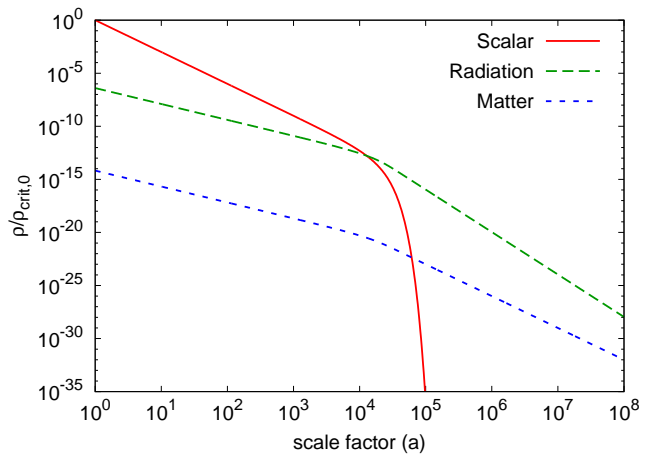}
 }
\caption{The evolution of the energy densities in the decaying scalar field and its decay products, as fractions of $\rho_\mathrm{crit,0}$: the critical density at $a=1$.  The scalar field is modeled as a pressureless fluid; its energy evolves as $\rho \propto a^{-3}$ until the expansion rate equals its decay rate at $a\simeq 10^{4}$.  While the Universe is scalar dominated, the scalar is feeding energy into the matter and radiation fields, and both the matter and radiation fields evolve as $\rho \propto a^{-3/2}$.  Once nearly all the scalar density is transferred to the matter and radiation fields, their energy densities evolve in the usual way: radiation as $\rho \propto a^{-4}$ and matter as $\rho \propto a^{-3}$. }
\label{fig:bkgrd}
\end{figure}

In the limit that $\gam t \ll1$, the Universe is scalar dominated, and the solutions to the background equations are
\beqa
\rhos &=& \rhos(t_0) \,a^{-3} \\
\rhor &=& \frac{3(1-f)}{5}[\rhos(t_0)\gam t_0 ]\,a^{-3/2} + c_r  \,a^{-4} \label{rhorsol}\\
\rhom &=& f [\rhos(t_0)\gam t_0 ]\,a^{-3/2} + c_m \,a^{-3}. \label{rhomsol}
\eeqa
In these equations, $c_r$ and $c_m$ are arbitrary constants to be determined by initial conditions.  If there is radiation or dark matter that does not originate from scalar decay, then $c_r$ and $c_m$ are positive.   If there was a time when there was no radiation or dark matter, then $c_r$ or $c_m$ are negative to make $\rhor=0$ or $\rhom=0$ at that time, as in Ref. \cite{GKR01}.  In either case, the contributions to the radiation and matter densities from the terms proportional to $c_r$ and $c_m$ become less significant as the scalar-dominated era continues.
We set our initial condition late in the scalar-dominated era, after any matter or radiation that did not originate from scalar decay is diluted and the system loses its memory of the beginning of scalar decay.
In this case, $c_r=c_m=0$, and the initial values (at $t=t_0$) of $\rhost, \rhort,$ and $\rhomt$ are determined by the chosen values for $\gamt$ and the branching ratio $f$.   During scalar domination, $H=2/(3t)$, which implies that $\rhos(t_0)\gam t_0 = (2/3)\gamt  \rhos(t_0)$.  It follows that
\beqa
\rhomt (t_0) &=& \frac{2}{3} f \gamt \rhost(t_0),\\
\rhort (t_0) &=& \frac{2}{5} (1-f)\gamt \rhost(t_0).
\eeqa
Finally, our assumption that the Universe is flat demands that $\rhost(t_0) +\rhort(t_0)+\rhomt(t_0) =1$.  We want the Universe to be initially dominated by the scalar energy density, so we must set $\gamt \ll 1$.  Figure \ref{fig:bkgrd} shows the evolution of the energy densities of the scalar, matter and radiation fields for $\gamt = 10^{-6}$ and $f = 10^{-8}$.  We see that the Universe becomes radiation dominated at $\adec \simeq \gamt^{-2/3}$. 

As described in Appendix \ref{sec:perts_derivation}, the evolution equations for perturbations in the scalar, radiation and matter fluids are derived by perturbing covariant versions of Eqs. (\ref{bkgdS})-(\ref{bkgdM}) \cite{KS84, MWU03, LM07}.
We work in conformal Newtonian gauge, with
\beq
\drm s^2 = a^2(\tau)[-(1+2\Psi)\drm \tau^2 + \delta_{ij}(1+2\Phi)\drm x^i \drm x^j].
\eeq
In addition to the metric perturbations $\Psi$ and $\Phi$, we must solve for the fractional density perturbations [\mbox{$\delta \equiv \delta\rho(\tau, \vec{x})/\rho^0(\tau)$}] in the scalar field ($\dels$), the radiation ($\delr$), and the matter ($\delm$).  The three fluids also have velocity perturbations [\mbox{$v^i \equiv \drm x^i/\drm\tau$}], and we solve for the 
divergences of the velocity fields for the scalar field ($\thes \equiv \grad\cdot \vec{v}_\phi$), the matter ($\them \equiv \grad\cdot \vec{v}_\mathrm{dm}$) and the radiation ($\ther \equiv \grad\cdot \vec{v}_r$).  The perturbation equations are
\begin{subequations} 
\label{pertset} 
\begin{align}
&\!\!\dot{\delta}_\phi+\thes + 3\dot{\Phi} = -a\gamo\Psi,\label{dels} \\
&\!\!\dot{\theta}_\phi + \scrih\thes+\nabla^2\Psi = 0, \label{thes} \\
&\!\!\dot{\delta}_r +\frac{4}{3}\ther+4\dot{\Phi} = (1-f)a\gamo\frac{\rhos^0}{\rhor^0}\left[\dels-\delr+\Psi\right], \label{delr}\\
&\!\!\dot{\theta}_r +\nabla^2\left(\frac{\delr}{4}+ \Psi\right) = (1-f)a \gamo \frac{\rhos^0}{\rhor^0}\left(\frac{3}{4}\thes-\ther\right), \label{ther}\\
&\!\!\dot{\delta}_\mathrm{dm}+\them + 3\dot{\Phi} = fa \gamo\frac{\rhos^0}{\rhom^0}\left[\dels-\delm+\Psi\right],\\
&\!\!\dot{\theta}_\mathrm{dm} + \nabla^2\Psi+\scrih\them = f a \gamo\frac{\rhos^0}{\rhom^0}\left[\thes-\them\right],
\end{align}
\end{subequations}
where a dot denotes differentiation with respect to conformal time $\tau$.   We also have the perturbed Einstein field equation
\beq
\nabla^2 \Phi +3\scrih\left(\scrih\Psi-\dot{\Phi}\right)=-4\pi G a^2\left(\rhos^0\dels+\rhor^0\delr+\rhom^0\delm\right).
\label{efe}
\eeq
Finally, we close the system of perturbation equations by noting that the absence of anisotropic stress implies $\Phi=-\Psi$.

To numerically solve these equations, we go to Fourier space and define the following dimensionless parameters: $E(a) \equiv H(a)/H_1$, $\tilde{k} \equiv k/H_1$, $\thest \equiv \thes/H_1$, $\thert \equiv \ther/H_1$, and $\themt \equiv \them/H_1$.  The equation set that we solve is given in Appendix \ref{sec:perts_ic}.  We use the scale factor $a$ as our time variable, and we set initial conditions when the Universe is scalar dominated and the mode is outside the Hubble horizon.  Since we start our solution when $a=1$ and $H=H_1$, these restrictions imply that $\gamt \ll 1$ and $\tilk \ll1$.  In Appendix \ref{sec:perts_ic} we derive the following initial conditions:
\begin{subequations} 
\label{ic} 
\beqa
\Phi(a_0) &=& \Phi_0 \\
\dels(a_0) &=& 2\Phi_0 + \frac{2}{3}\tilk^2\Phi_0a_0\\
\delr(a_0) &=& \Phi_0+\frac{46}{63}\tilk^2\Phi_0a_0 \\
\delm(a_0) &=& \Phi_0+\frac{2}{3}\tilk^2\Phi_0a_0 \\
\thest(a_0) =\thert(a_0)&=&\themt(a_0)= -\frac{2}{3}\tilk^2\Phi\sqrt{a_0}, 
\eeqa
\end{subequations}
with $a_0 = 1$.  These initial conditions neglect terms that are ${\cal O}(\tilk^4 a_0)$ and ${\cal O}(\gamt a_0)$.

The initial conditions presented above differ from the characteristics of superhorizon adiabatic perturbations in a radiation-dominated universe, for which $\delr = 2\Phi$ and $\delm = (3/4)\delr$.   Since the matter and radiation originated from a single scalar field, we should be left with adiabatic perturbations in the matter and radiation after the scalar vanishes.  Indeed, modes that are superhorizon at the time of reheating [$\tilk < \adec E(\adec)]$ evolve when $a\simeq \adec$, and this evolution takes the perturbations from the initial values given above to the conditions for adiabatic superhorizon modes in a radiation-dominated universe.  The gravitational potential transitions from its initial value $\Phi_0\equiv\Phi(t_0)$ to $(10/9)\Phi_0$ as the Universe transitions from being effectively matter dominated to being radiation dominated.  The radiation density perturbation grows from $\delr(t_0)=\Phi_0$ to $2(10/9)\Phi_0$.   Meanwhile, the matter density perturbation grows from $\delm(t_0)=\Phi_0$ to $(5/3)\Phi_0$.  Thus, after the Universe becomes radiation dominated, we have $\delm = (3/4)\delr$ on superhorizon scales.  Once these modes enter the horizon, they behave exactly as expected for adiabatic perturbations in a radiation-dominated universe, with $\Phi = (10/9)\Phi_0$ on superhorizon scales.  In the following sections, we will consider the smaller-scale modes that enter the horizon while the Universe is still scalar dominated.

\section{Perturbations in the Radiation Fluid}
\label{sec:radpert}
We briefly review the standard evolution of a perfect-fluid radiation perturbation in a radiation-dominated universe, and then compare this to the evolution of radiation perturbations when followed through reheating.  

\begin{figure}[tb]
 \centering
 \resizebox{3.5in}{!}
 {
      \includegraphics{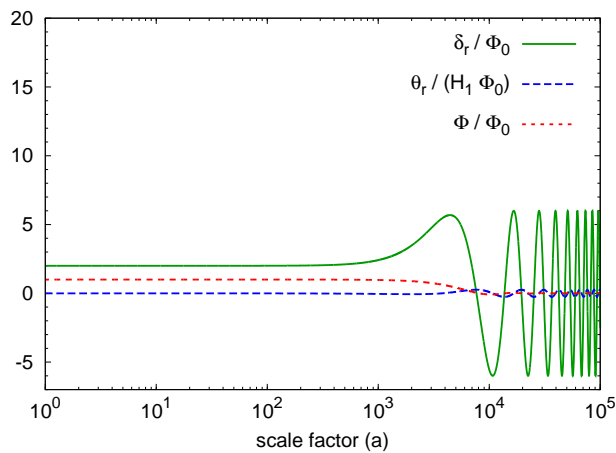}
 }
\caption{Evolution of $\delr/\Phi_0$ , $\thert/\Phi_0$, and $\Phi/\Phi_0$ for a mode that enters the horizon at $a\simeq1000$.  In this cosmological scenario, the Universe is radiation dominated; there is no era of scalar domination.}
\label{Fig:RFT}
\end{figure}

In a radiation-dominated universe, the equations describing the evolution of perfect-fluid radiation perturbations can be solved exactly and take the form
\begin{widetext} 
\begin{subequations} 
\label{RFTeqns} 
\begin{eqnarray}
\Phi(\chi) &=& \frac{9\Phi_0}{\chi^3}\left[\sqrt{3} \sin \left(\frac{\chi}{\sqrt{3}}\right) - \chi \cos \left(\frac{\chi}{\sqrt{3}}\right)\right],\\
\delr(\chi) &=&  \frac{6\Phi_0}{\chi^3}\left[2\sqrt{3}(\chi^2-3)\sin \left(\frac{\chi}{\sqrt{3}}\right) - \chi (\chi^2-6) \cos \left(\frac{\chi}{\sqrt{3}}\right) \right],\\
\thert(\chi) &=& - \frac{3\sqrt{3} \tilk\Phi_0}{2\chi^2}\left[2\sqrt{3}\chi \cos \left(\frac{\chi}{\sqrt{3}}\right) + (\chi^2-6) \sin \left(\frac{\chi}{\sqrt{3}}\right)\right], \label{RFTtheta}
\end{eqnarray}
\end{subequations}
\end{widetext}
where $\Phi_0$ is the initial value for $\Phi$ set when the mode is far outside the horizon \cite{Bertschinger06}, and $\chi \equiv k \tau = \tilk a/[a_{\mathrm{RD}}^2 E(a_{\mathrm{RD}})]$.  Here, $a_{\mathrm{RD}}$ denotes a value of the scale factor well after reheating (deep into radiation domination) and $E(a_{\mathrm{RD}})$ is the dimensionless Hubble factor as discussed in Sec.~\ref{sec:threefluid}.    In Fig.~\ref{Fig:RFT} we show the evolution of the radiation perturbations in the standard scenario.   The perturbations are frozen when a given mode is superhorizon and oscillate after the mode enters the horizon.   As the amplitude of oscillations in $\Phi$ decay as  $A_\Phi \propto \chi^{-2}$, the equations for the perfect-fluid variables $\delr$ and $\thert$ reduce to those of a harmonic oscillator.  After they are well in the horizon the perturbations in $\delr$ and $\thert$ quickly asymptote to cosine and (out-of-phase) sine solutions with amplitudes $A_{\delr} = 6 \Phi_0$ and $A_{\thert} =(\tilk\sqrt{3}/4)A_{\delr}$ respectively.

We now compare the subhorizon behavior of the radiation perturbations following a period of reheating to the standard behavior given in Eq. (\ref{RFTeqns}).  To make this comparison, we must take care to compare modes that have the same physical wave vector at some temperature $T_{\mathrm{RD}}$ long after the reheating process is complete.
These modes have not typically spent the same amount of time inside the Hubble horizon.  If a mode enters the horizon prior to reheating, then it does so earlier than it would have if the Universe never had an epoch of early scalar domination.  For instance, the mode shown in Fig.~\ref{Fig:RFT} enters the horizon at $a=1000$ if the Universe was always radiation dominated.  But if the Universe was scalar dominated up to $a\simeq10^4$, then this mode enters the horizon when $a=100$.  In general, for modes that enter the horizon during scalar domination, the value of $a$ at horizon entry in the scalar-dominated scenario is $\kdec/k$ times the value it would have been in the standard scenario, where $\kdec = \adec H(\adec)$ is the wave number of the mode that enters the horizon at reheating.

\begin{figure}
 \centering
 \resizebox{3.5in}{!}
 {
      \includegraphics{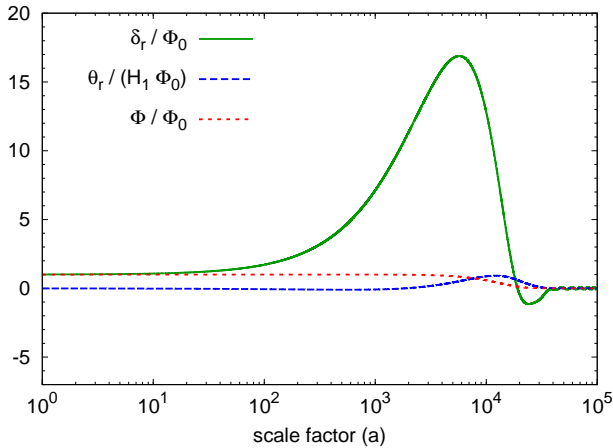}
 }
\caption{Evolution of $\delr/\Phi_0$, $\thert/\Phi_0$, and $\Phi/\Phi_0$ for a mode ($\tilk = 0.1$) that enters the horizon at $a\simeq100$.  In this cosmological scenario, the Universe was scalar dominated for $a\lsim10^4$, and then it became radiation dominated.  When the Universe is radiation dominated, $\delr$ oscillates with an amplitude of $0.085\Phi_0$.  For this mode, \mbox{$k/\kdec \simeq 11$}.}
\label{Fig:smallmode}
\end{figure}
\begin{figure}
 \centering
 \resizebox{3.5in}{!}
 {
      \includegraphics{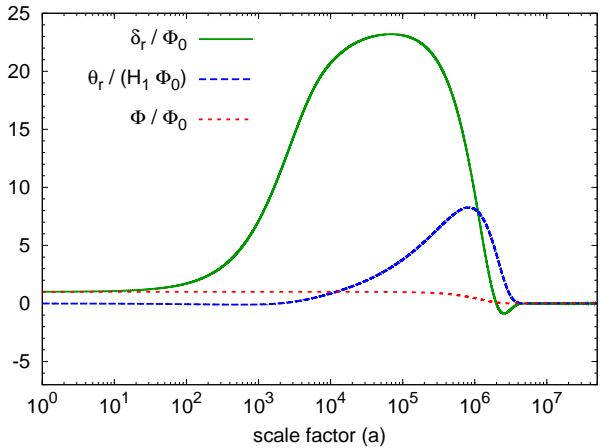}
 }
\caption{Evolution of $\delr/\Phi_0$, $\thert/\Phi_0$, and $\Phi/\Phi_0$ for a mode ($\tilk = 0.1$) that enters the horizon at $a\simeq100$.  In this cosmological scenario, the Universe was scalar dominated for $a\lsim10^6$, and then it became radiation dominated.  When the Universe is radiation dominated, $\delr$ oscillates with an amplitude of $0.0007\Phi_0$.  For this mode, \mbox{$k/\kdec \simeq 114$}.}
\label{Fig:smallmode2}
\end{figure}

Figure~\ref{Fig:smallmode} shows the same mode as Fig.~\ref{Fig:RFT} in a universe that is scalar dominated for $a\lsim 10^4$ (corresponding to the choice $\gamt = 10^{-6}$).   The evolution of this mode is drastically different than the standard evolution shown in Fig.~\ref{Fig:RFT}.  Initially, $\delr$ grows considerably while the Universe is scalar dominated.  However, after reheating this enhancement is lost and $\delr$ oscillates with a small amplitude.  The late-time oscillations have the same period as the oscillations given by Eq.~(\ref{RFTeqns}), but the amplitude of the fluid perturbations is suppressed by a factor of $\sim\!\!\!70.6$ with respect to the standard case, with $A_{\delr}\simeq (6/70.6)\Phi_0 \simeq 0.085 \Phi_0$
and $A_{\thert} = (\tilk\sqrt{3}/4) A_{\delr}$.
The $\Phi$ solution does not begin to decay until after reheating in this scenario but subsequently decays as $\chi^{-2}$ with a similarly suppressed amplitude relative to the standard evolution.
Figure~\ref{Fig:smallmode2} shows the $\tilk=0.1$ mode in a universe with $\gamt = 10^{-9}$.  The late-time solution exhibits similar behavior, only now with an amplitude suppressed by $\sim\!\!\!7600$ with respect to the standard case, so that   $A_{\delr}\simeq (6/7600)\Phi_0 \simeq 0.0007 \Phi_0$.
In short, the radiation perturbation at late times follows, up to a phase shift, Eq.~(\ref{RFTeqns}) with a suppressed amplitude that depends on the value of $k/\kdec$.
For $k/\kdec\gsim 20$ we find $A_{\delr} \lesssim 10^{-3}\Phi_0$ with the precise value modulating by an order of magnitude as a function of $k/\kdec$.  For $20\gsim k/\kdec \gsim 3$ $A_{\delr}$ increases with decreasing $k$, reaching a peak of $A_{\delr}\simeq 10 \Phi_0$.  For smaller values of $k$,  $A_{\delr}$ decreases as $k$ decreases and flattens out at a value of $A_{\delr} \simeq (60/9)\Phi_0$ for $k/\kdec\lsim 0.1$.  Modes with $k/\kdec \lesssim 0.1$ are superhorizon at the time of reheating and thus their evolution follows Eq.~(\ref{RFTeqns}) with $\Phi_0 \rightarrow (10/9)\Phi_0$ since in this scenario $\Phi_0$ is defined during a period of matter domination.

This behavior can be understood by re-examining Eqs.~(\ref{delr}) and (\ref{ther}).
During the scalar-dominated epoch, $\dot{\Phi}=0$, $\dels \propto a$, and $\dels$ grows large.  Since the scalar is decaying to radiation, this large scalar overdensity 
becomes a source for $\delr$, and $\delr$ grows considerably.  The increase in the radiation overdensity $\delr$ in turn sources a higher $\ther$, representing an outflow of radiation fluid in real space, through Eq.~(\ref{ther}). This outflow slows the growth in $\delr$, which eventually saturates and has $\dot{\delta}_r=0$. Meanwhile, as the decay continues, $\ther$ keeps growing and the outflow balances the source terms from scalar decay in Eq.~(\ref{delr}).  Eventually, the scalar source term becomes subdominant, and $\delr$ decreases as the outflow continues.  This decrease in $\delr$ slows the outflow, and $\ther$ decreases to a point where $\delr$ and $\ther$ begin to oscillate out of phase.  Due to the relatively large value of $\dels$, these oscillations are forced; the equilibrium point is shifted to positive $\delr$, and the first trough in $\delr$ is very shallow.  However, by the time $\delr$ emerges from the first underdensity, the scalar density is zero and henceforth $\delr$ undergoes free oscillations with a small amplitude.

In summary, the radiation perturbation long after reheating is described by Eq.~(\ref{RFTeqns}), but with $\Phi_0$ replaced by $T(k)\Phi_0$ and $a \rightarrow a + a_0$, where $a_0$ is the phase shift resulting from the fact that oscillations started at reheating rather than horizon entry.  For $k \gsim 20 \kdec$, $T(k) \lsim 10^{-3}$, and the perturbations are suppressed.   As $k$ decreases, $T(k)$ increases, reaching $T(k) \simeq 1.5$ for $2\lsim k/\kdec \lsim 4$.  For smaller values of $k$, $T(k)$ again decreases until it levels out at $T(k/\kdec \lsim 0.1) = 10/9$.  The major effect of a period of scalar domination on the radiation perturbation is a suppression of power on scales that entered the horizon during the scalar-dominated era.  These scales are very small as $k_{\mathrm{RH}}$ is given by
\beqa
\kdec &=& 1690 \, \left(\frac{T_\mathrm{RH}}{100 \, \mbox{GeV}}\right)\left(\frac{100}{g_{*S}}\right)^{1/3}\left(\frac{g_*}{100}\right)^{1/2} \, \mbox{pc}^{-1} \nonumber\\
&=& 0.0117\left(\frac{T_\mathrm{RH}}{1 \, \mbox{MeV}}\right)\left(\frac{10.75}{g_{*S}}\right)^{1/3} \left(\frac{g_*}{10.75}\right)^{1/2}  \, \mbox{pc}^{-1} \nonumber
\eeqa
where $T_\mathrm{RH}$ is the temperature of the radiation bath when the Universe becomes radiation dominated.  In this expression,  $g_* \equiv \rho_r/[(\pi^2/30)T^4]$, and both $g_*$ and $g_{*S}$ are evaluated at $T_\mathrm{RH}$.  
If the dark matter is a thermal relic that decoupled from the radiation bath after reheating, perturbations with $k > \kdec$ will be suppressed. 

This suppression of density perturbations is in principle relevant for any theory in which dark matter has origins in the thermal bath after reheating, such as the standard WIMP paradigm.  
The characteristic kinetic decoupling temperature $T_\mathrm{KD}$ in WIMP models is necessarily below the chemical decoupling temperature and thus the reheat temperature $T_\mathrm{RH}$ if they are
thermally produced after reheating \cite{PSK2006}.    However, in models where kinetic decoupling occurs immediately after chemical decoupling, the
nominal cutoff from kinetic decoupling and the cutoff from altered perturbation evolution in this reheating scenario can be comparable and a more detailed analysis including both effects 
is warranted.   In models where the dark matter is never in equilibrium with the standard model but is nevertheless produced from a thermal plasma after reheating, such as in 
feebly interacting massive particle (FIMP) models where the dark matter abundance freezes-in \cite{HJMW2010}, in hylogenic models \cite{DMST2010}, in some hidden-sector models \cite{S2009,DS2010},
and generally for models where $T_\mathrm{KD} \gtrsim T_\mathrm{RH}$, the reheating cutoff is expected to be the dominant cutoff in the small-scale matter power spectrum.

\section{Perturbations in the Matter Fluid}
\label{sec:matpert}
\begin{figure}
 \centering
 \resizebox{3.5in}{!}
 {
      \includegraphics{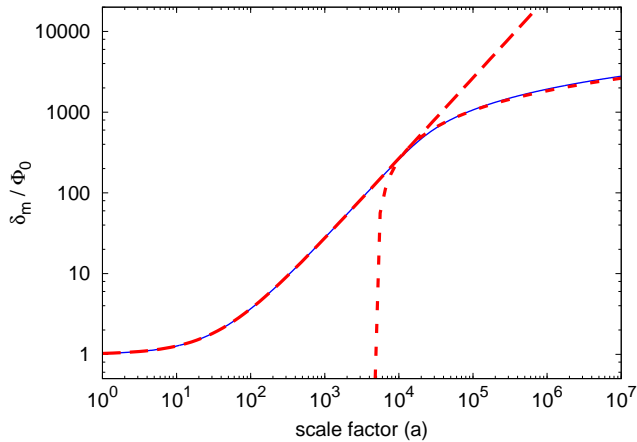}
 }
\caption{The evolution of the fractional density perturbation in the matter field ($\delm$).  This mode enters the horizon when $a=25$, which is during the scalar-dominated era.  During scalar-domination, subhorizon matter perturbations grow linearly with $a$.  The Universe becomes radiation dominated when $a=10^{4}$, at which point $\delm$ grows logarithmically. The long-dashed curve depicts Eq.~(\ref{linmod}), which describes the evolution of $\delm$ during scalar domination.  The short-dashed curve depicts Eq.~(\ref{mdmodel}) with $\adec = 1.29\gamt^{-2/3}$, which fits the evolution of $\delm$ after reheating.}
\label{Fig:deltaMevol}
\end{figure}

Since the radiation perturbation does not retain any of its growth during the scalar-dominated era, an epoch of scalar domination results in primordial structure growth 
only if the dark matter decouples from the radiation before the Universe becomes radiation dominated.  We now consider the evolution of the decay-produced matter perturbation 
$\delm$ for modes that enter the horizon while the Universe is dominated by the oscillating scalar field.
For these modes, $\delm$ grows linearly with $a$ between horizon entry and reheating, and then $\delm$ grows logarithmically after the Universe becomes radiation dominated.  An example is shown in Fig.~\ref{Fig:deltaMevol}, which shows the evolution of $\delm$ for $\tilk = 0.2$ and $\gamt = 10^{-6}$; this mode enters the horizon when $a=25$, and the Universe becomes radiation dominated when $a=10^4$.

We now derive an analytic model for the evolution of $\delm$.  Prior to the decay of the scalar field, the Universe is effectively matter dominated; $a^2 E(a) = \sqrt{a}$ and $\Phi$ remains constant after the mode enters the Hubble horizon.  The solutions derived in Appendix \ref{sec:perts_ic} for the velocities $\thest(a)$ and $\themt(a)$ hold until the Universe becomes radiation dominated.  It follows that
\beq
\delm^\prime(a) = \frac{2}{3} \tilk^2 \Phi_0 + \frac{3}{2a}\left[\dels - \delm -\Phi \right].
\eeq
The quantity in the square brackets is initially zero, and it remains zero until the Universe becomes radiation dominated, as one can verify by considering $\dot{\delta}_\phi-\dot{\delta}_\mathrm{dm}$.  Therefore, during scalar domination,
\beq
\delm(a) = \Phi_0 + \frac{2}{3}\tilk^2 \Phi_0a,
\label{linmod}
\eeq
and we see that the early-time solution derived in Appendix \ref{sec:perts_ic} holds after the mode enters the Hubble horizon.  This model for $\delm(a)$ is depicted in Fig.~\ref{Fig:deltaMevol}, and we see that it matches the numerical solution for $a\lsim \adec$.

The linear growth begins when the linear term exceeds the initial value [$(2/3)\tilk^2 a \gsim 1$].  We define $a_\mathrm{lin} \equiv 3/(2\tilk^2)$.  For modes that enter the horizon during scalar domination, $a_\mathrm{hor} = \tilk^{-2}$, so $a_\mathrm{lin}$ is larger than $a_\mathrm{hor}$.  
Since \mbox{$\delm = \Phi_0$} prior to the onset of linear growth, at the time of scalar decay \mbox{$\delm(\adec) = (\adec/a_\mathrm{lin}) \Phi_0 = (2/3) \tilk^2 \adec \Phi_0$}.  After the Universe becomes radiation dominated, $\delm(a)$ grows logarithmically, and $a \delm^\prime (a)$ is constant.  We can therefore set $a \delm^\prime (a) = \adec \delm^\prime (\adec) = (2/3) \tilk^2 \Phi_0\adec$.
It follows that, for modes with $k/\kdec \gsim 1$,
\beq
\delm(a) = \frac{2}{3} \adec \tilk^2 \Phi_0 \left[1+\ln \left(\frac{a}{\adec}\right)\right]
\label{mdmodel}
\eeq
after reheating.  This model for $\delm(a)$ is depicted in Fig.~\ref{Fig:deltaMevol}, and we see that it matches the numerical solution for $a\gsim \adec$.

\subsection{The Matter Transfer Function}
\label{sec:tk}
Due to the superhorizon evolution of $\Phi$ during the transition from scalar domination to radiation domination, the definition of the transfer function is ambiguous.  It is standard to define the transfer function $T(k)$ through the relation
\beq
\delm(k, a\gg a_\mathrm{eq}) = \frac{3}{5} \frac{k^2}{\Omega_\mathrm{M} H_0^2} \Phi_\mathrm{p}(k) T(k) D(a),
\label{Tdef}
\eeq
where $H_0$ is the present-day Hubble parameter, $\Omega_\mathrm{M}$ is the current density  of matter divided by the current critical density, $\Phi_\mathrm{p}$ is the initial potential fluctuation, and $D(a)$ is the growth function normalized to make $D(a)=a$ during matter domination.  With this definition, $T(k) =1$ on large scales because modes that enter the horizon after matter-radiation equality have $\Phi = (9/10)\Phi_\mathrm{p}$ at horizon entry.  In our scenario, however, these modes enter the horizon with $\Phi = \Phi_0$, where $\Phi_0$ is the superhorizon potential perturbation during scalar domination.  To define our transfer function, we will use the standard definition above, but we note that  $\Phi_\mathrm{p}$ is the superhorizon potential perturbation during radiation domination; $\Phi_\mathrm{p} = (10/9)\Phi_0.$ 

\begin{figure}
 \centering
 \resizebox{3.5in}{!}
 {
      \includegraphics{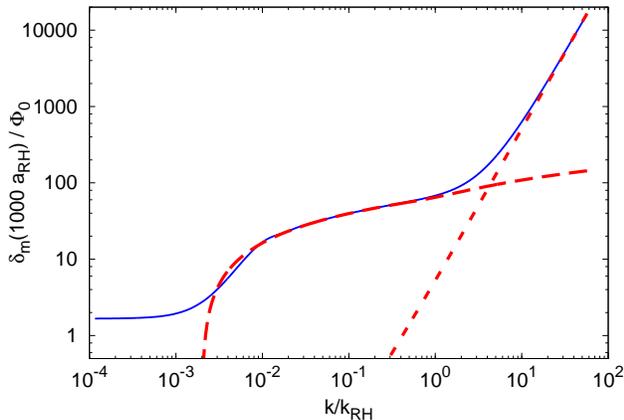}
 }
\caption{The fractional perturbation in the matter density evaluated after the scalar decays \mbox{$(a = 1000 \gamt^{-2/3})$}.  The solid curve is the numerical evaluation.  Modes with $k/\kdec \leq 0.001$ are still superhorizon at this value of the scale factor.  The long-dashed curve is a plot of Eq.~(\ref{logfunc}) and describes subhorizon modes that enter the horizon after the scalar decays.  The short-dashed line is a plot of Eq.~(\ref{mddelk}) and describes modes that enter the horizon during scalar domination.}
\label{Fig:deltaM(a)}
\end{figure}

Figure \ref{Fig:deltaM(a)} shows how $\delm$ at a set value of the scale factor depends on $k$.  In the cosmology depicted in this figure, $\gamt = 10^{-6}$, which implies that $\adec = 10^4$ and $\kdect \equiv\adec E(\adec) = 0.0088$.  The matter perturbation is numerically evaluated at $a=10^7$ for a wide range of $\tilk$ values; this is the solid curve in Fig.~\ref{Fig:deltaM(a)}.  We see that the function $\delm(1000\adec, k/\kdec)$ contains three different behaviors.  First, for $k/\kdec \lsim 0.001$, we have $\delm(1000\adec, k/\kdec) = 5/3 \Phi_0$.  These modes have not yet entered the horizon at $a=10^{7}$, and so they are fixed at the adiabatic ``initial" condition, as described in Section \ref{sec:threefluid}.

Modes with $0.001 \lsim k/\kdec \lsim 1$ enter the horizon during radiation domination.  These modes grow logarithmically once they enter the horizon; as long as \mbox{$\delta \rhor \gg \delta \rhom$}, their behavior is well-described by the function
\beq
\delm(a) = \frac{10}{9} \Phi_0 \left[A \ln \left(\frac{B \,a}{a_\mathrm{hor}}\right)\right]
\label{logfunc}
\eeq
with $A=9.11$ and $B = 0.594$, as given in Ref.~\cite{HS96}.  The long-dashed curve in  Fig.~\ref{Fig:deltaM(a)} shows this function with $a = 1000\adec$ and $a_\mathrm{hor} E(a_\mathrm{hor}) \equiv \tilk$.  We see that it is an excellent fit for $0.01 \lsim k/\kdec\lsim 1$.  We conclude that the standard transfer function can be used for modes with $k/\kdec\lsim 1$; these modes are unaffected by the era of scalar domination.  

Modes with $k/\kdec \gsim 1$ enter the horizon while the Universe is dominated by the oscillating scalar field. To obtain the transfer function for these modes, we express $\delm(a)$ in the same form as Eq.~(\ref{logfunc}) and then extract the values for $A$ and $B$.  From Eq.~(\ref{mdmodel}), we see that $A = (3/5)\adec\tilk^2$ and $\ln B = 1+ \ln (a_\mathrm{hor}/\adec)$.  We now evaluate $A$ and $B$ in terms of $k/\kdec$.  First we note that we never used $\adec = \gamt^{-2/3}$ in the derivation of Eq.~(\ref{mdmodel}).  Rather, $\adec$ was used as the transition point between linear and logarithmic growth.  We find that Eq.~(\ref{mdmodel}) fits the numerical solution for $\delm(a)$ if $\adec = 1.29 \gamt^{-2/3}$, as shown in Fig.~\ref{Fig:deltaMevol}.  We also find that $\gamt^{2/3} \simeq 1.29 \kdect^2$ for a wide range of values of $\gamt$.  We conclude that we should use $\adec = \kdect^{-2}$ when evaluating $A$ and $B$.  Since these modes enter the horizon deep in the era of scalar domination, we also have $a_\mathrm{hor} = \tilk^{-2}$.  We conclude that 
\beqa
\delm(a) &=& \frac{10}{9} \Phi_0 \left[A \ln \left(\frac{B \,a}{a_\mathrm{hor}}\right)\right] \label{mddelk}\\
A &=& \frac{3}{5}\left(\frac{k}{\kdec}\right)^2 \nonumber\\
B &=& \frac{e \, \kdec^2}{k^2}\nonumber
\eeqa
for modes with $k/\kdec \gsim 1$ evaluated after scalar decay while \mbox{$\delta \rhor \gg \delta \rhom$}.  This model, with $a = 10^7$ and $a_\mathrm{hor} E(a_\mathrm{hor}) \equiv \tilk$, is shown by the short-dashed line in Fig.~\ref{Fig:deltaM(a)} and we see that it is an excellent fit to the numerical solution when $k/\kdec \gsim 10$.  To smoothly connect this solution to the $k\lsim\kdec$ solution ($A=9.11$ and $B=0.594$), we found functions $A(k/\kdec)$ and $B(k/\kdec)$ that fit the numerical solution for $\delm(a,k)$ when they are inserted into Eq.~(\ref{mddelk}).  These functions are given in Appendix \ref{sec:ABfit}.  

Now that we have $A(k)$ and $B(k)$, we can obtain the behavior of $\delm(a)$ during the matter-dominated era by matching Eq.~(\ref{mddelk}) to the decaying and growing modes of the Meszaros equation, which is valid when \mbox{$\delta \rhom \gg \delta \rhor$} \cite{Meszaros74}.  While the baryons are still coupled to the photons, they do not fall into the potential wells created by the dark matter density perturbations.  Prior to baryon decoupling and after matter-radiation equality, $\delm(a)$ is given by \cite{HS96}
\beq
\delm =\frac{3A}{2}f_1 \left(\frac{10}{9} \Phi_0\right) \ln \left[\left(\frac{4}{e^3}\right)^\frac{f_2}{f_1}\frac{B a_\mathrm{eq}}{a_\mathrm{hor}}\right]{\cal D}\left(a\right), \label{latedelm}
\eeq
where $f_1$, $f_2$ and $\alpha$ are determined by the baryon fraction $f_\mathrm{b} \equiv \rho_\mathrm{b} /(\rho_\mathrm{b}+\rhom)$,
\beqa
f_1 &=& 1 - 0.568 f_\mathrm{b} + 0.094 f_\mathrm{b}^2 \nonumber\\
f_2 &=& 1 - 1.156 f_\mathrm{b} + 0.149 f_\mathrm{b}^2 - 0.074 f_\mathrm{b}^3 \nonumber,
\eeqa
and ${\cal D}(a)$ is the growing solution to the Meszaros equation.  If baryons do not participate in gravitational collapse, 
\beq
{\cal D}(a) = \left(1+\frac{a}{a_\mathrm{eq}}\right)^{-\alpha} {_2F_1}\left[\alpha, \alpha+\frac{1}{2}; 2\alpha+\frac{1}{2}; \frac{a_\mathrm{eq}}{a+a_\mathrm{eq}}\right],
\label{fullD}
\eeq
where ${_2F_1}[a,b;c;x]$ is Gauss's hypergeometric function, and 
\beq
\alpha = \frac{1}{4}\left[1-\sqrt{1+24(1-f_\mathrm{b})}\right].
\eeq
Long after matter-radiation equality, Eq.~(\ref{fullD}) reduces to ${\cal D}(a) = (a/a_\mathrm{eq})^{-\alpha}$; dark matter overdensities experience slower than linear growth if the baryons do not fall into their potential wells.  Consequently, matter perturbations on scales smaller than the baryon Jeans length grow slower than larger-scale perturbations \cite{Bertschinger06}, and we will need to account for this suppression in our transfer function.

To complete our derivation of the matter transfer function, we evaluate $a_\mathrm{eq}/a_\mathrm{hor}$ for modes that enter the horizon during scalar domination:
\beq
\frac{a_\mathrm{eq}}{a_\mathrm{hor}} = \frac{\sqrt{2} k}{k_\mathrm{eq}} \left(\frac{k}{\kdec}\right)\left(\frac{g_{*S}[T_\mathrm{RH}]}{3.91}\right)^{2/3}\left(\frac{3.36}{g_*[T_\mathrm{RH}]}\right)^{1/2}.
\label{aeqahor}
\eeq 
The terms containing $g_*$ and $g_{*S}$ account for changes in the number of relativistic species between reheating and matter-radiation equality, and the product of these terms is always between 1.1 and 1.6.  The product $B{a_\mathrm{eq}}/{a_\mathrm{hor}}$ is proportional to ${\kdec}/{k_\mathrm{eq}}$, and it appears in Eq.~(\ref{latedelm}) logarithmically.  Since ${\kdec}/{k_\mathrm{eq}}\gg1$, the terms containing $g_*$ and $g_{*S}$ have a negligible effect on $\delm$, and we do not consider them further.   For modes that enter the horizon during radiation domination, $a_\mathrm{eq}/a_\mathrm{hor} = \sqrt{2} k/k_\mathrm{eq}$, and in general, we find that
\beq
\frac{a_\mathrm{eq}}{a_\mathrm{hor}} = \frac{\sqrt{2} k}{k_\mathrm{eq}}\left[1+\left(\frac{k}{k_\mathrm{RH}}\right)^{4.235}\right]^{1/4.235}.
\label{ahorfit}
\eeq
fits the numerical solution to $k = a_\mathrm{hor} H(a_\mathrm{hor})$ for all values of $k$.

Inserting Eq.~(\ref{aeqahor}) into Eq.~(\ref{latedelm}) and evaluating the transfer function without baryons gives
\beq
T(k \gg \kdec)  = \frac{3}{4} \left(\frac{k_\mathrm{eq}}{\kdec}\right)^2\ln \left[\left(\frac{4}{e^3}\right)\frac{e \sqrt{2}\kdec}{\keq}\right].
\label{Tbd}
\eeq
Thus we see that $T(k \gg \kdec)$ is scale-invariant. These modes grow linearly with the scale factor between horizon entry and scalar decay, and they grow linearly with the scalar factor after the Universe becomes matter dominated.  The transfer function characterizes deviations from linear growth in $\delm(a)$.  During radiation domination, these modes grew logarithmically instead of linearly, and so the transfer function depends only on the duration of the radiation-dominated era: \mbox{$T(k \gsim \kdec) \sim (\kdec/k_\mathrm{eq})^{-2}\ln [\kdec/k_\mathrm{eq}]$}.  Given current measurements \mbox{($k_\mathrm{eq}$ = 0.0098 Mpc$^{-1}$)} \cite{wmap7param}, we note that
\beqa
\frac{\kdec}{k_\mathrm{eq}} &=& 1.72\times10^{11} \, \left(\frac{T_\mathrm{RH}}{100 \, \mbox{GeV}}\right)\left(\frac{100}{g_{*S}}\right)^{1/3}\left(\frac{g_*}{100}\right)^{1/2}, \nonumber\\
&=& 1.18\times10^6\left(\frac{T_\mathrm{RH}}{1 \, \mbox{MeV}}\right)\left(\frac{10.75}{g_{*S}}\right)^{1/3}\left(\frac{g_*}{10.75}\right)^{1/2}, \nonumber
\eeqa
where $g_*$ and $g_{*S}$ are evaluated at $T_\mathrm{RH}$.   

When we consider structure formation in Section \ref{sec:structure}, we will need a transfer function that is applicable to all $k$ values, including $k\lsim k_\mathrm{eq}$, and includes the effects of baryons and neutrinos on structure formation.  We use CAMB Sources \cite{CAMB} to compute the matter transfer function for $k/\keq \leq 8.2 \times 10^5$.  On very small scales ($k/\keq \gsim 5\times10^5$), the transfer function computed by CAMB ($T_\mathrm{CAMB}$) has the same scale dependence as the transfer function ($T_\mathrm{EH}$) provided by Eisenstein \& Hu (1998) \cite{EH98}, so we extend the transfer function to larger $k$ by taking 
\beq
T\left({k}/{\keq}\geq8.2 \times 10^5\right) =T_\mathrm{EH}(k) \frac{T_\mathrm{CAMB}(k/\keq=8.2 \times 10^5)}{T_\mathrm{EH}(k/\keq=8.2 \times 10^5)}.
\eeq
When evaluating the transfer function, we take \mbox{$H_0=70.4$} km/s/Mpc, $\Omega_\mathrm{M} = 0.272$, and $\Omega_\mathrm{b} = 0.0456$, following Ref.~\cite{wmap7param}.  

Even after recombination, the baryons have nonzero pressure, and they do not participate in gravitational collapse on scales that are smaller than the baryon Jeans length $1/k_\mathrm{Jb}$ \cite{Bertschinger06}.  While matter perturbations on scales with $k < k_\mathrm{Jb}$ grow linearly with the scale factor after recombination, the growth of smaller-scale perturbations still follows ${\cal D}(a)$, given by Eq.~(\ref{fullD}).  We account for this suppression by using CAMB Sources to determine the density perturbation at a redshift of 50, and then we use Eq.~(\ref{Tdef}) to find $T(k)D(z=50)$, which we refer to as the transfer function at $z=50$.  We then apply a scale-dependent growth function, normalized to unity at $z=50$, to obtain the density perturbation at other redshifts.  At $z=50$, $T_\mathrm{CAMB}(k) \simeq T_\mathrm{EH}(k)$ for $k/\keq \lsim 10^4$, but $T_\mathrm{CAMB}/T_\mathrm{EH}$ decreases as $k$ increases until it reaches a new plateau: $T_\mathrm{CAMB}(k) \simeq 0.77 T_\mathrm{EH}(k)$ for $k/\keq \gsim 5 \times 10^5$.  This suppression results from the slower growth of these small-scale perturbations after recombination, which is not included in $ T_\mathrm{EH}(k)$. 

We base our scale-dependent growth function on the ratio $T_\mathrm{CAMB}/T_\mathrm{EH}$ at $z=50$;  for $z\gsim 3$, modes with $k/\keq \gsim 10^5$ are proportional to ${\cal D}(a)$, while modes with $k/\keq \lsim 10^4$ are proportional to \mbox{$(2/3)+(a/a_\mathrm{eq})$}.  Intermediate scales smoothly interpolate between these two values; since we will primarily be concerned with modes with $k/\keq \geq 10^7$, the details of the transition are unimportant.  At later times, the cosmological constant suppresses perturbation growth, and we assume that all scales follow the standard growth function $D(a)$ for $z \lsim 3$.  An explicit expression for our growth function is given in Appendix \ref{sec:ABfit}.

\begin{figure}
 \centering
 \resizebox{3.5in}{!}
 {
      \includegraphics{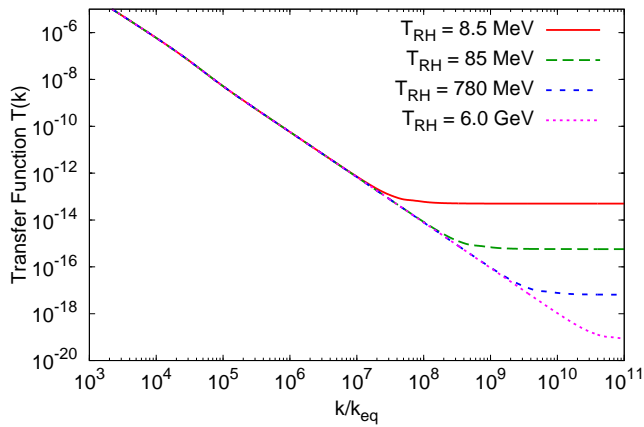}
 }
\caption{The matter transfer function in the absence of free-streaming effects for four values of the reheat temperature: $T_\mathrm{RH}=8.5 \mbox{ MeV}, 85 \mbox{ MeV}, 780 \mbox{ MeV}$, and $6.0$ GeV, corresponding to $\kdec/\keq = 10^7,10^8,10^9$ and $10^{10}$, respectively.  Perturbation modes that enter the horizon while the Universe is radiation dominated have $T(k) \propto \ln[k]/k^2$, but modes that enter the horizon prior to reheating have a constant-valued $T(k)$.}
\label{Fig:Transk}
\end{figure}

To modify the transfer function to include a period of scalar domination, we multiply it by the ratio $\delm(T_\mathrm{RH})/\delm(T_\mathrm{RH}=\infty)$, where $\delm$ is evaluated after matter-radiation equality and before baryon decoupling.  From Eq.~(\ref{latedelm}), we see that, for modes with $k\gg\keq$, including a period of scalar domination takes $T(k)\rightarrow R(k)T(k)$, where
\beq
R(k) = \frac{A(k) \ln \left[\left(\frac{4}{e^3}\right)^\frac{f_2}{f_1}\frac{B(k) a_\mathrm{eq}}{a_\mathrm{hor}(k)}\right]}{9.11 \ln \left[\left(\frac{4}{e^3}\right)^\frac{f_2}{f_1} 0.594  \frac{\sqrt{2} k}{k_\mathrm{eq}}\right]}.
\eeq
In this expression, $a_\mathrm{hor}(k)$ is given by Eq.~(\ref{ahorfit}),  and $A(k)$ and $B(k)$ are given by the fitting functions in Appendix \ref{sec:ABfit} for $k \geq 0.05 \kdec$.  For $k \leq 0.05 \kdec$, $\delm$ is not affected by the period of scalar domination, as seen in Fig.~\ref{Fig:deltaM(a)}.  Therefore, we take $R(k \leq 0.05 \kdec) = 1$.  The resulting transfer functions (evaluated at $z=0$) are shown in Fig.~(\ref{Fig:Transk}) for four values of the reheat temperature.  As expected from Fig.~\ref{Fig:deltaM(a)}, we see that the transfer function deviates from its standard value when $k$ exceeds $\kdec$, and it is scale-invariant for $k \gsim 10\kdec$.

\subsection{Free-Streaming Effects}
\label{sec:freestream}
We have thus far assumed that the dark matter has no momentum when it is produced from the decay of the scalar field.  We now relax that assumption and introduce $\vdec$: the average velocity ($\drm \vec{x}/\drm \tau$) of the dark matter particles when the Universe became radiation dominated.  We will continue to assume that the dark matter is always non-relativistic, so $\vdec \ll 1$.  

If the dark matter particles have momentum, then their free-streaming will tend to erase structures on scales smaller than the free-streaming horizon $\lfsh$ \cite{GHS05, Bertschinger06, BLRV09}:
\beq
\lfsh(t) = \int_{t_\mathrm{RH}}^t \frac{\langle v\rangle}{a} \, \drm t,
\eeq
where $\langle v\rangle = \vdec (\adec/a)$ is the average velocity of the dark matter particles after reheating.  The free-streaming horizon does not change significantly after matter-radiation equality, so we may neglect dark energy and take
\beq
\lfsh(a) = \frac{\vdec \adec}{H_0\sqrt{\Omega_\mathrm{r}}} \int_{\adec}^a \frac{\drm a^\prime}{a^\prime\sqrt{1+a^\prime/a_\mathrm{eq}}},
\eeq
where $\Omega_\mathrm{r}$ is the present-day radiation density divided by the critical density.  This integral can be evaluated analytically \cite{BLRV09}:
\beq
\lfsh(a) = \frac{2\vdec \adec}{H_0\sqrt{\Omega_\mathrm{r}}} \left[\sinh^{-1} \sqrt{\frac{a_\mathrm{eq}}{\adec}}-\sinh^{-1}\sqrt{\frac{a_\mathrm{eq}}{a}}\right].
\label{lfsh}
\eeq

The effects of free-streaming on the matter power spectrum can be approximated by introducing a Gaussian cutoff to the transfer function:
\beq
T(k) = \exp\left[-\frac{k^2}{2\kfsh^2}\right]T_0(k),
\label{Tfs}
\eeq
where $T_0(k)$ is the transfer function without free-streaming calculated in the previous section, and \mbox{$\kfsh = \lfsh^{-1}$} \cite{GHS04, GHS05, LZ05, Bertschinger06}.
If $\kdec/\kfsh > 1$, then the free-steaming cutoff in the transfer function will exponentially suppress the scale-invariant portion of $T_0(k)$ given by Eq.~(\ref{Tbd}).  The growth of density perturbations during scalar domination will be completely erased by the free-streaming of the dark matter particles after reheating.  The ratio $\kdec/\kfsh$ depends only weakly on the reheat temperature; if we neglect changes in the number of relativistic species after reheating,
\beq
\frac{\kdec}{\kfsh} = {2\vdec}{\left[\sinh^{-1} \sqrt{\frac{\kdec\sqrt{2}}{k_\mathrm{eq}}}-\sinh^{-1}\sqrt{\frac{a_\mathrm{eq}}{a}}\right]}.
\eeq
For a wide range of reheating temperatures ($10 \mbox{ Mev} \lsim T_\mathrm{RH} \lsim 10 \mbox{ GeV}$), $\kdec/\kfsh \simeq \vdec/0.06$.

If the dark matter produced via decay does not interact with the radiation fluid, then a low $\vdec$ will occur if the mass of the dark matter (or more generally the total mass of the decay products) is nearly degenerate in mass with the decaying scalar.
If the dark matter instead interacts with the plasma through a higher-dimensional effective operator (such as a four-fermion operator), then we generically expect a larger interaction cross section at higher energies.  In this case, we expect dark matter particles born with large momentum to rapidly lose energy to the plasma until interactions become inefficient and the diffusion length to be far shorter than the free-streaming length would be without interactions.  Calculating the evolution of dark matter perturbations including the thermalization and diffusion of high-energy dark matter particles goes beyond the scope of this work, but after interactions become inefficient we expect the formalism developed here will describe subsequent perturbation evolution with $\vdec \sim \sqrt{E_{\rm KD}/m_{\rm DM}}$ set by the energy at which particles typically kinetically decouple from the plasma.  Finally, we mention that if some fraction or all of the dark matter did not originate from scalar decay and was present during the scalar-dominated era, then these particles likely have low velocity \cite{GG08}, and we expect that they would experience the same perturbation growth described at the beginning of Section~\ref{sec:matpert}.

\section{Primordial Structures}
\label{sec:structure}

We have shown that matter perturbations on scales smaller than the horizon size at reheating and larger than the matter particle's free-streaming horizon are enhanced relative to their amplitudes in a universe that was always radiation dominated.  We will now use the Press-Schechter formalism \cite{PS74}
to analyze the ramifications this enhancement has on the formation of dark matter halos.  We must first compute the rms density perturbation in a sphere containing an average mass $M$:
\beq
\sigma^2(M,z) = \int \frac{\drm^3k}{(2\pi)^3} \left[D_{50}(k,z)T_{50}(k)\right]^2 P_\mathrm{p}(k) F^2(kR),
\label{sigmadef}
\eeq
where $T_{50}(k)$ is the transfer function evaluated \mbox{$z=50$}; $D_{50}(z,k)$ is the scale-dependent growth function described in Appendix \ref{sec:ABfit}; $P_\mathrm{p}(k)$ is the power spectrum of superhorizon density perturbations during radiation domination; and $F(kR)$ is a filter function that suppresses contributions from modes with \mbox{$k^{-1}\ll R = [3M/(4\pi \rho_{\mathrm{m},0})]^{1/3}$}, where $\rho_{\mathrm{m},0}$ is the present-day matter density.  We determine $P_\mathrm{p}(k)$ by setting the power spectrum of superhorizon curvature fluctuations $\Delta^2_{\cal R}(k) = 2.44\times10^{-9} \times (k/k_0)^{n-1}$, where \mbox{$k_0 = 0.002\mbox{ Mpc}^{-1}$} and $n$ is the scalar spectral index: \mbox{$n=0.963\pm0.012$ \cite{wmap7param}}.  

It is customary to use a spherical top-hat window function in real space as a filter when computing $\sigma(M)$; this choice implies $F(kR) = 3[\sin(kR)-(kR)\cos(kR)]/(kR)^3$.  However, this filter function is problematic when the effects of a scalar-dominated era are considered.  We showed in Section \ref{sec:tk} that $T(k)$ is scale invariant for $k\gsim \kdec$.  Since $P_\mathrm{p}(k) \propto k^n$, the integrand in Eq.~(\ref{sigmadef}) for $k\gsim \kdec$ is proportional to $k^{n+2} F^2(kR) \drm k$.  With the top-hat filter function, $F^2(kR) \simeq \cos^2(kR)/(kR)^4$ for $k\gg R^{-1}$, so for $kR \gg 1$ and $k\gsim \kdec$, the integrand in Eq.~(\ref{sigmadef}) is proportional to $k^{n-2} \drm k$.  Since $n$ is only slightly less than unity, this integral is nearly divergent, and the portion of the integral with $k\gg 1/R$ makes a significant contribution to $\sigma^2(M)$.  In short, the standard top-hat filter fails to prevent modes with $k^{-1}\ll R$ from contributing to $\sigma^2(M)$ when $T(k)$ is scale-invariant for large $k$.  

To solve this problem, we use a different filter function when calculating $\sigma^2(M)$; we convolve the spherical top-hat window function with a Gaussian window function with a much smaller radius.  The resulting filter function is
\beqa
F(kR) &=& \exp\left[-\frac{1}{2}k^2(\alpha R)^2\right] \nonumber\\
&&\times\frac{3}{(kR)^3}\left[\sin(kR)-(kR)\cos(kR)\right]
\eeqa
with $\alpha\ll1$.  For $\alpha \lsim 10^{-4}$, this window function is virtually indistinguishable from a top-hat of radius $R$ in real space, and $\sigma(M)$ for $R\gg1/\kdec$ changes by less than 0.01\% compared to its value with $\alpha = 0$.  This robustness prompts us to set $\alpha =10^{-4}$ when computing $\sigma(M)$.  We note, however, that the variance on smaller scales is more sensitive to the choice for $\alpha$; for $R\ll1/\kdec$, changing $\alpha$ by a factor of ten in either direction changes $\sigma(M)$ by about ten percent.  

\begin{figure}
 \centering
 \resizebox{3.5in}{!}
 {
      \includegraphics{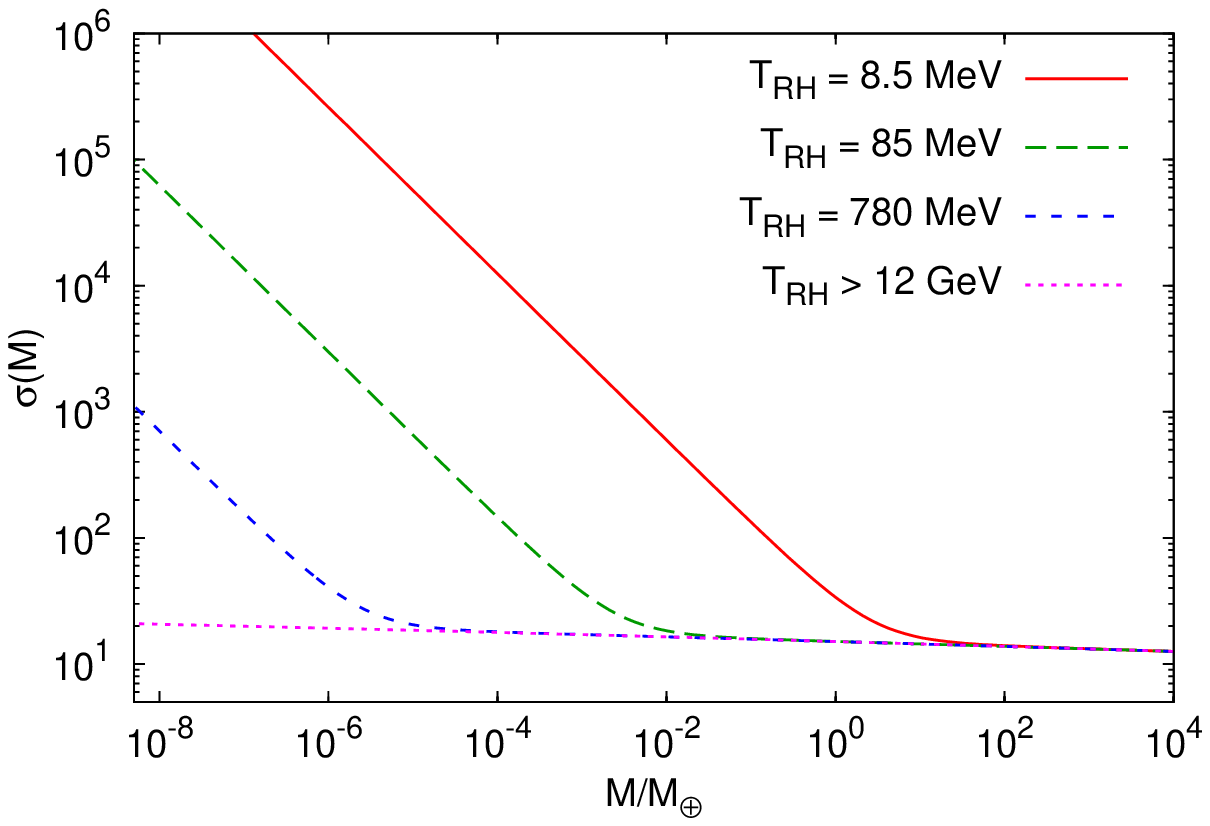}
 }
\caption{The present-day rms density perturbation $\sigma$ in a sphere containing an average mass $M$ for three values of the reheat temperature: \mbox{$T_\mathrm{RH}=8.5 \mbox{ MeV}, 85 \mbox{ MeV}$ and $780 \mbox{ MeV}$}.  For these values of $T_\mathrm{RH}$, $\mdec = 54 \,M_\oplus, 0.054\,M_\oplus$, and $5.4\times10^{-5}\,M_\oplus$.  We also show $\sigma(M)$ for reheat temperatures greater than 12 GeV; in the mass range shown here, the presence of a scalar-dominated era does not affect $\sigma(M)$ if \mbox{$T_\mathrm{RH}>12$ GeV}.  Note that $\sigma(M\gsim\mdec)$ depends only weakly on $M$ and is independent of  $T_\mathrm{RH}$, while $\sigma(M\lsim\mdec)$ is significantly enhanced compared to $\sigma$ for higher values of $T_\mathrm{RH}$.  Also, for all values of $T_\mathrm{RH}$, \mbox{$\sigma(M\lsim\mdec)\propto M^{-(n+3)/6}$}, where $n=0.963$ is the scalar spectral index.}
\label{Fig:SigTrh}
\end{figure}

The filter function ensures that $\sigma(M)$ is sensitive to scales that enter the horizon during scalar domination only if $R(M)\lsim\kdec^{-1}$.  Consequently, it is useful to define a characteristic mass $\mdec$ such that \mbox{$R(\mdec)\equiv \kdec^{-1}$}:
\beq
\mdec = 32.7 M_\oplus \,\left(\frac{10 \, \mbox{MeV}}{T_\mathrm{RH}}\right)^3\left(\frac{g_{*S}[T_\mathrm{RH}]}{10.75}\right)\left(\frac{10.75}{g_*[T_\mathrm{RH}]}\right)^{3/2}.
\eeq
For $M>\mdec$, the filter function effectively restricts the integral in Eq.~(\ref{sigmadef}) to $k\lsim\kdec$, and $\sigma(M)$ is insensitive to the reheat temperature.   In contrast, if $M<\mdec$, $\sigma(M)$ is most sensitive to scales with $k>\kdec$.  The transfer function is scale-invariant at these scales, so the density power spectrum is a simple power law $P(k)\propto k^n$.  It follows that $\sigma(M\lsim\mdec)\propto M^{-(n+3)/6}$, as shown in Fig.~\ref{Fig:SigTrh}.  

\begin{figure}
 \centering
 \resizebox{3.5in}{!}
 {
      \includegraphics{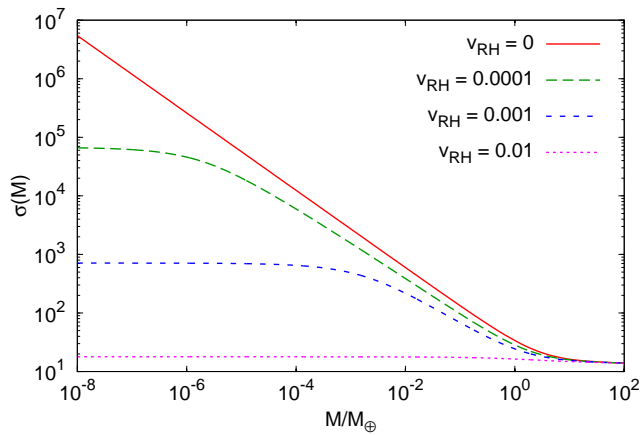}
 }
\caption{The rms density perturbation $\sigma$ in a sphere containing an average mass $M$ for four values of the mean dark matter particle velocity at reheating $(v_\mathrm{RH})$, in units where $c=1$.  The reheat temperature is 8.5 MeV.  If the dark matter particles have a nonzero velocity dispersion, free-streaming erases density perturbations on scales smaller than the free-streaming horizon, making $\sigma(M)$ nearly constant for small masses.}
\label{Fig:SigVdec}
\end{figure}

As discussed in Section \ref{sec:freestream}, we can include the effects of the dark matter particles' random motions by adding a Gaussian cutoff to the transfer function, as in Eq.~(\ref{Tfs}).  Since the resulting transfer function is no longer scale-invariant for $k>\kfsh$, we expect $\sigma(M\lsim\mdec)$ to deviate from the power law behavior shown in Fig.~\ref{Fig:SigTrh} when $M\lsim \mfsh$, where \mbox{$R(\mfsh)\equiv \kfsh^{-1}= \lfsh$}.   Figure \ref{Fig:SigVdec} shows that this is the case; in this figure, $T_\mathrm{RH}$ is 8.5 MeV, and $\sigma(M)$ is shown for four values of $\vdec$.  For $\vdec = 0.01, 0.001$, and $0.0001$, Eq.~(\ref{lfsh}) implies \mbox{$\mfsh = 0.31\, M_\oplus, 3.1\times10^{-4} M_\oplus$}, and $3.1\times10^{-7}M_\oplus$, respectively, and we see that $\sigma(M)$ is nearly constant for $M\lsim\mfsh$.   Figure \ref{Fig:SigVdec} also shows that the growth of the matter density perturbations prior to reheating is completely erased by free-streaming if $\vdec \gsim 0.01$; this is also true for other values of $T_\mathrm{RH}$.

We have seen that the growth of matter perturbations during the scalar-dominated era leads to an enhancement $\sigma(M)$ for $M\lsim\mdec$, provided that $\vdec \lsim 0.01$.  Since $\mdec \lsim 1000 \, M_\oplus$, the amplitude of $\sigma(M\lsim \mdec)$ depends on the amplitude of the scalar power spectrum on scales that are much smaller than the pivot scale \mbox{$k_0 = 0.002$ Mpc$^{-1}$}.  Consequently, $\sigma(M\lsim \mdec)$ is very sensitive to changes in the scalar spectral index $n$.  We can quantify the dependence of $\sigma(M)$ on $n$ by noting that
\beq
\frac{\sigma(M, n_1)}{\sigma(M, n_2)} \simeq \left[\frac{1}{k_0 R_\sigma(M)}\right]^{\frac{n_1-n_2}{2}},
\eeq
where $R_\sigma(M)$ is the average scale that contributes significantly to $\sigma(M)$.  For $M\gsim \mdec$, $R_\sigma(M) = R(M)$ because the transfer function and filter function suppress the contribution from smaller scales. 
For $M\lsim \mdec$ and $\vdec = 0$, all scales between $\alpha R$ and $R$ contribute equally, so $R_\sigma(M) \simeq R(M)\sqrt{\alpha}$.  Free-streaming introduces an additional small-scale cutoff; for \mbox{$\alpha R \lsim \lfsh \lsim R$}, $R_\sigma(M)\simeq \sqrt{R(M)\lfsh}$, and $R_\sigma(M)\simeq \lfsh$ if $R\lsim\lfsh$.  For example, if $T_\mathrm{RH} = 8.5$ MeV and $\vdec =0$, $\sigma(0.001\, \mdec)$ increases by 37\% if $n$ is 0.987 instead of 0.963.  If we instead take $\vdec =0.001$, then $\sigma(0.001\, \mdec)$ increases by 32\% for the same change in $n$; the introduction of a free-streaming cutoff has reduced effects of increasing the scalar spectral index.

Now that we have calculated $\sigma(M)$, we can use the Press-Schechter formalism \cite{PS74} to obtain a halo mass function:
\beq
\frac{d n}{d \ln M} = \sqrt{\frac{2}{\pi}}\,\frac{\rho_{\mathrm{m},0}}{M} \left|{\frac{d \ln\sigma}{d \ln M}}\right| \frac{\delcoll}{\sigma(M,z)} \exp\left[-\frac{\delcoll^2}{2\sigma^2(M,z)}\right],
\label{PSmf}
\eeq
where $n$ is the comoving number density of halos with mass $M$ at redshift $z$, and $\delcoll$ is the critical linear overdensity.  For $z\gsim 2$, $\delcoll= 1.686$, and it decreases slightly at smaller redshift.   Since an era of scalar domination changes $\sigma(M)$ for $M\lsim\mdec$, only the abundance of these small-mass halos will be affected.  These small-mass halos form at very high redshift, and then they merge to form larger-mass halos.  The Press-Schechter mass function does not account for subhalos, however, so it will be most useful for us to examine the halo mass function at high redshift, before halos with $M\lsim\mdec$ are absorbed into larger halos.  Rather than consider the number density of these objects, it is more enlightening to compute the fraction of the mass that is contained in these objects as a function of redshift:
\beq
\frac{d f}{d \ln M} = \frac{M}{\rho_{\mathrm{m},0}} \frac{d n}{d \ln M} .
\label{massfracdef}
\eeq

\begin{figure}
 \centering
 \resizebox{3.5in}{!}
 {
      \includegraphics{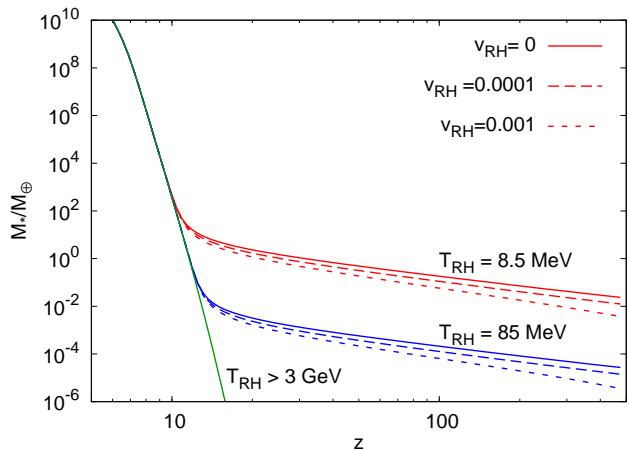}
 }
\caption{The value of $M_*$ such that $\sigma(M_*)=\delcoll$ plotted as a function of redshift for different values of the reheat temperature $T_\mathrm{RH}$ and the average particle velocity at reheating, $\vdec$, in units where $c=1$.  For $T_\mathrm{RH} =8.5$ MeV and $T_\mathrm{RH} =85$ MeV, the reheat horizon mass is $\mdec = 54 \,M_\oplus$ and $0.054\, M_\oplus$, respectively.  We see that $M_*(z)$ is a power law for $M_*\lsim \mdec$.}
\label{Fig:mstar}
\end{figure}

The ratio $\delcoll/\sigma$ plays a pivotal role in the Press-Schechter mass function, and it is useful to define a function $M_*(z)$ such that $\sigma(M_*,z)=\delcoll$.    Figure \ref{Fig:mstar} shows $M_*(z)$ for different values of the reheat temperature and $\vdec$.  We see that $M_*(z)$ decreases precipitously with increasing redshift for $M\gsim \mdec$.  For smaller masses, $M_*(z)$ transitions to a power law.  Free-streaming by the dark matter particles steepens this power law slightly because $\sigma(M)$ increases less quickly with decreasing $M$ if $\vdec$ is nonzero, as shown in Fig.~\ref{Fig:SigVdec}.  Also recall that $\sigma(M)$ is nearly constant for $M\lsim \mfsh$; this constancy would cause $M_*(z)$ to decrease rapidly with increasing redshift if $M_*\lsim \mfsh$.  We see in Fig.~\ref{Fig:mstar}, however, that if $\vdec\lsim 0.001$, $M_*\gsim \mfsh$ for all $z<500$.  Therefore, we expect that free-streaming will not prevent structures from growing at redshifts $z\lsim500$, provided that $\vdec\lsim0.001$. 

Before we consider the effects of free-streaming on the halo mass function further, we examine the mass function with $\vdec=0$.   If $\sigma(M)$ is a power law, as it is for $M\ll \mdec$, then the mass function depends on redshift only through the ratio $M/M_*(z)$.  In this case, the differential bound mass fraction $df/d\ln M$ peaks at $M=M_*$; for $d \ln \sigma/d\ln M = -(n+3)/6$, the maximum value is \mbox{$df/d\ln M|_{M_*}=0.32$}, and 31\% of the dark matter is contained in halos with $M_*\leq M\leq 5M_*$.   Since the abundance of halos with $M\gsim M_*$ is exponentially suppressed, a negligible fraction of the mass is contained in halos with $M>5M_*$.  Provided that $5M_* \ll \mdec$, the fact that $\sigma(M)$ is not a power law for $M\gsim\mdec$ will not affect the bound fraction.

\begin{figure}
 \centering
 \resizebox{3.5in}{!}
 {
      \includegraphics{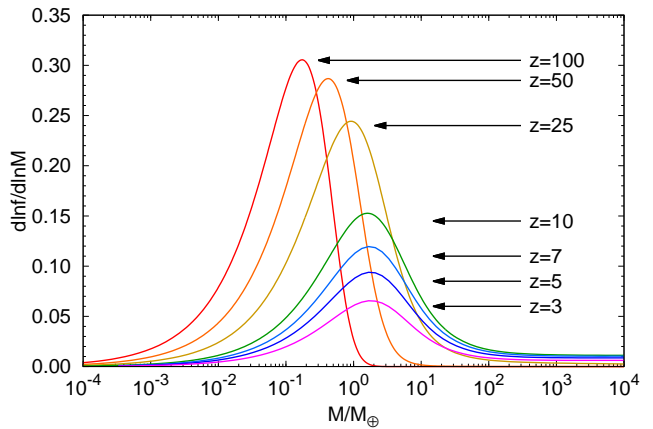}
 }
\caption{The differential fraction of the dark matter mass that is bound into halos of mass $M$ [see Eq.~(\ref{massfracdef})] plotted at several redshifts for $T_\mathrm{RH} = 8.5$ MeV.  While $z\gsim 11$, the critical mass $M_*$ is smaller than the reheating horizon mass $\mdec$, and the peak moves to larger masses as the redshift decreases.  When $z\leq10$, $M_*>\mdec$, and the peak remains fixed at $2M_\oplus$.  The peak decreases in amplitude because the fraction of mass contained in these microhalos decreases as they are absorbed into larger halos.}
\label{Fig:massfrac}
\end{figure}

The $5M_* \ll \mdec$ case is exemplified by the $z=100$ curve in Fig.~\ref{Fig:massfrac}, which shows the differential bound fraction \mbox{$df/d\ln M$} for $T_\mathrm{RH}=8.5$ MeV.   For this reheat temperature, $\mdec = 54 M_\oplus$, and Fig.~\ref{Fig:mstar} shows that $M_* = 0.18 M_\oplus$ at $z=100$.  Integrating the $z=100$ curve in Fig.~\ref{Fig:massfrac} reveals that almost half of the dark matter is contained in halos with $M>0.1M_\oplus$ at this redshift.  In the standard cosmology, the fraction of the dark matter contained in such halos is only $10^{-10}$ at $z=100$.  As the redshift increases past 100, the differential bound fraction will keep the same shape as the $z=100$ curve while the peak slides to smaller masses.  The other curves in Fig.~\ref{Fig:massfrac} show what happens to the bound mass function as the redshift decreases.  While $z\gsim 11$,  $M_*> \mdec$, and the peak in \mbox{$df/d\ln M$} follows $M_*$, moving to larger masses as the redshift decreases.  The peak height decreases because \mbox{$|d\ln \sigma/d\ln M|_{M_*}$} is decreasing, as seen in Fig.~\ref{Fig:SigTrh}.

\begin{figure}
 \centering
 \resizebox{3.5in}{!}
 {
      \includegraphics{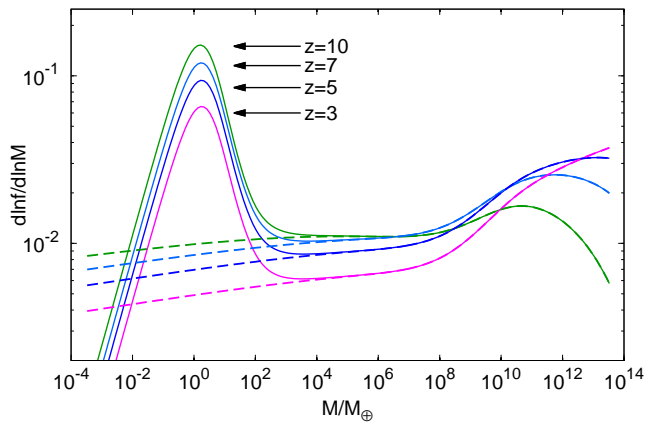}
 }
\caption{The differential fraction of the dark matter mass that is bound into halos of mass $M$ plotted at several redshifts.  The solid curves show the bound fraction for \mbox{$T_\mathrm{RH}=8.5$ MeV}, while the dashed curves show the bound fraction in the absence of a scalar-dominated era.  The two scenarios are indistinguishable for $M\gsim1000\, M_\oplus$.  The abundance of halos is higher for halos with $M\gsim 10^{10} M_\oplus$ because baryons fall into these halos, making them grow faster.}
\label{Fig:massfraclog}
\end{figure}

When $z\lsim11$, $M_*>\mdec$ for $T_\mathrm{RH}=8.5$ MeV, and larger-mass halos begin to form.  These larger halos absorb some of the microhalos formed at higher redshift, and so the fraction of mass in these microhalos decreases, as seen in Fig.~\ref{Fig:massfrac}.  The transfer of mass from small halos to larger halos is more apparent in Fig.~\ref{Fig:massfraclog}, which also shows the differential bound fraction without a scalar-dominated era.  We see that a period of scalar domination does not affect the abundance of halos with $M\gg \mdec$.  Without a scalar-dominated era, however, halos with $M \lsim \mdec$ are far less abundant and they form much later.  
For example, if $T_\mathrm{RH}=8.5$ MeV, the abundance of Earth-mass microhalos peaks at a redshift of $z=32$ when $df/d\ln M|_{M_\oplus}=0.25$.  In contrast, isolated Earth-mass microhalos are most abundant at a redshift of $z=12$ in the standard scenario, and even then $df/d\ln M|_{M_\oplus}=0.010$.

\begin{figure}
 \centering
 \resizebox{3.5in}{!}
 {
      \includegraphics{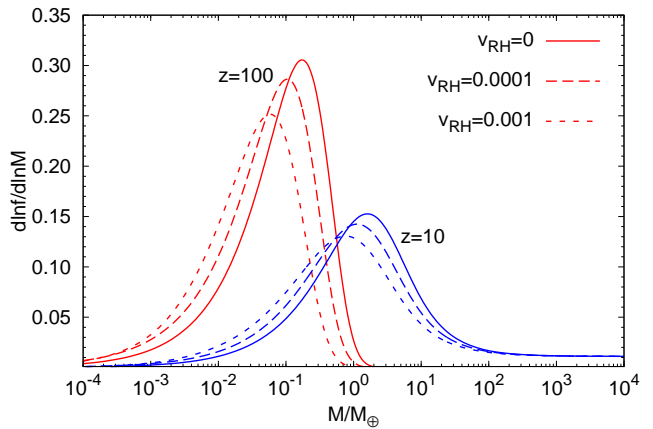}
 }
\caption{The differential fraction of the dark matter mass that is bound into halos of mass $M$ for different values of the mean dark matter particle velocity at reheating, as a fraction of the speed of light ($v_\mathrm{RH}$).  The reheat temperature is $T_\mathrm{RH} = 8.5$ MeV, and the function is shown at two redshifts.  We see that free-streaming moves the peak in the bound mass fraction to smaller masses and decreases its amplitude.}
\label{Fig:massfracvdec}
\end{figure}

Given our understanding of the bound mass fraction when $\vdec=0$, we can now examine the effects of free-streaming.  Figure~\ref{Fig:mstar} shows that free-streaming decreases $M_*$ when $M_*\lsim \mdec$, and Fig.~\ref{Fig:SigVdec} shows that free-streaming decreases $d \ln \sigma/d\ln M$ when $M\lsim \mdec$.  We conclude that free-streaming will shift the peak in the bound mass fraction to smaller halo masses and decrease its amplitude.  Figure~\ref{Fig:massfracvdec} confirms that this is indeed the case.  We see that free-streaming delays the formation of microhalos of a given mass and suppresses their abundance; 
for example, if $\vdec=0.001$ and $T_\mathrm{RH}=8.5$ MeV, the abundance of Earth-mass microhalos peaks at a redshift of $z= 22$ when $df/d\ln M|_{M_\oplus}=0.17$.  This is still significantly higher than the standard abundance of these microhalos.  We conclude that free-streaming does not erase the structural imprints of an extended scalar-dominated era, provided that $\vdec\lsim 0.001$.   Microhalos with $M<\mdec$ still form earlier and in far greater numbers than they would in the standard scenario, but free-streaming does slightly delay their formation and suppress their abundances.

\section{Summary and Discussion}
\label{sec:summary}
The content of the Universe prior to the onset of big bang nucleosynthesis is unknown.  It is possible that the inflaton continued to dominate the energy density of the Universe long after  inflation ended.  After inflation, the inflaton oscillates around the minimum of its potential, and if the potential is quadratic around its minimum, then the time-averaged pressure of the scalar field vanishes.  It is also possible that the inflaton decayed quickly to radiation, and then a second oscillating scalar field may have come to dominate the Universe, as in some versions of the curvaton scenario \cite{Mollerach90, LM97, LW02, MT01} and in several realizations of string theory \cite{CCQR93, BKN94, AKK10}.  Finally, the inflaton may have decayed into a short-lived heavy particle that dominated the energy density of the Universe prior to its decay.   In all of these scenarios, the Universe was effectively matter dominated prior to nucleosynthesis. 

During this early matter-dominated era, subhorizon perturbations in the dominant component of the Universe grow linearly with the scale factor \cite{KMZ85, JK08, JLM10a, EFG10}.  We have investigated what happens to these density perturbations after the dominant component during the early matter-dominated era decays and the Universe becomes radiation dominated.  We model the reheating of the Universe using three fluid components: a pressureless ``scalar" that decays into a tightly-coupled radiation plasma and decoupled nonrelativistic dark matter.  These dark matter particles may have a small, but nonzero, velocity dispersion when they are created, and we examine the effects of free-streaming on their perturbations.  
 
We find that, although subhorizon radiation perturbations grow during the scalar-dominated era due to the coupling between the radiation and the scalar field, this growth is erased after the Universe becomes radiation dominated.  When the scalar field energy density vanishes, these radiation perturbations begin to oscillate, but the amplitude of these oscillations is much smaller than the maximum value of the radiation density perturbation during the scalar-dominated era.  Moreover, the amplitude of the oscillations in the radiation density perturbations for modes that enter the horizon well before reheating is significantly smaller than the amplitude of modes that enter the horizon after the Universe becomes radiation dominated.  An early era of matter domination suppresses perturbations in the radiation density field on scales smaller than the horizon scale at reheating. 
If the dominant form of dark matter is \emph{not} created directly from reheating but instead from the radiation bath after reheating, this radiation suppression results in a suppression of the matter power spectrum for $k \gtrsim  23.4\, (T_{\mathrm{RH}}/ 2 \,{\rm MeV})\, \,{\rm kpc}^{-1}$.
Depending on the details of the dark-matter microphysics, this reheating suppression may be the primary cutoff in the matter power spectrum.  This could include corners of WIMP parameter space with inefficient kinetic coupling \cite{PSK2006}, FIMP models \cite{HJMW2010}, and hylogenic models \cite{DMST2010}.

The effects of reheating on dark matter produced directly from the decay of the scalar can be much more dramatic.
Unlike the radiation density perturbation, the perturbation in this dark matter density is enhanced by the growth it experiences during the scalar-dominated era.  While the Universe is scalar dominated, dark matter density perturbations grow linearly with the scale factor after they enter the Hubble horizon.  When the Universe becomes radiation dominated, the dark matter perturbations grow logarithmically from their amplitude at the moment of reheating.  Consequently, perturbations on scales that enter the horizon prior to reheating are significantly enhanced compared to larger-scale perturbations, and the matter power spectrum on small scales is proportional to $k^n$, where $n$ is the scalar spectral index.  Free-streaming breaks this scale invariance and exponentially suppresses the power spectrum on scales smaller than the free-streaming horizon.  We found that free-streaming completely erases the perturbations that grow prior to reheating if the average velocity dispersion of the dark matter particles at reheating is greater than $0.01c$.

We used the Press-Schechter halo mass function \cite{PS74} to investigate how the growth of perturbations prior to reheating affects the formation of dark matter halos.   As expected, only halos with masses less than the horizon mass at reheating are impacted.  The reheating horizon mass ($\mdec$) is proportional to $T_\mathrm{RH}^{-3}$ and equals $260\,M_\oplus$ if the reheat temperature was $T_\mathrm{RH} = 5$ MeV.  We found that microhalos with $M<\mdec$ form much earlier than they would if the reheat temperature were higher, and they contain a much larger fraction of the dark matter.  
For instance, if $T_\mathrm{RH} = 8.5 $ MeV, then the Press-Schechter mass function predicts that 15\%, 43\% and 65\% of the dark matter is contained in microhalos with masses greater than $10^{-6}\, M_\odot$ at redshifts of 100, 50, and 25, respectively.  In contrast, the corresponding fractions are $10^{-10}$, 0.04\% and 5\% if the reheat temperature is higher than 100 MeV.  Numerical simulations of the formation of the first microhalos confirm this prediction; they find that only 1.5\% of the dark matter is in bound in these microhalos at a redshift of 31 \cite{IME10}, and 5\% of the dark matter is in these microhalos at a redshift of 26 \cite{DMS05}.

We conclude that low-temperature reheating with direct dark matter production results in an abundance of earth-mass or smaller microhalos.  Prior to the formation of larger dark matter halos, these microhalos contain a significant fraction of the dark matter.  What happens to these microhalos as they merge to form larger halos, and do they survive until the present day in the Milky Way's halo?  The fate of the earth-mass microhalos that form in the standard cosmological scenario has been studied extensively \cite{DMS05, ZHATS07, BDE06, GG07, GGMDS07, AZ07, SKM10, IME10}; these subhalos survive their absorption into larger halos, but then they lose most of their mass due to interactions with stars.  The survival probability for a microhalo near the Sun has been calculated to be as high as 0.17 \cite{BDE06}, but this factor depends strongly on the microhalo's orbit, and some studies predict that nearly all the local microhalos are either destroyed by stellar encounters \cite{ZHATS07, GG07, AZ07} or lose a significant fraction of their mass \cite{GGMDS07, SKM10}.  In the low-temperature reheating scenarios considered here, however, these microhalos are far more numerous than in the standard scenario; even if there is a high probability that they do not survive inside the Galaxy, there may still be a sizable population in our stellar neighborhood.  

Furthermore, the microhalos in the low-temperature-reheating scenario form far earlier than their standard counterparts, and consequently, they are far denser.  The virial density of a microhalo is proportional to the critical density at the time of its formation; at high redshifts, the mean density of a microhalo that formed at a redshift $z_f$ is therefore proportional to $(1+z_f)^3$.   Ref. \cite{Berezinsky10} found that the microhalo survival probability near the Sun increases sharply as the mean density within the clump increases; while a microhalo that formed at a redshift of 65 has a survival probability of 0.17 at the Sun's location, a microhalo that formed at a redshift of 120 has survival probability of 0.5, and a microhalo that formed at a redshift of 300 has survival probability of 0.9.
It is therefore reasonable to expect that many of the microhalos produced in the low-temperature-reheating scenario are intact subhalos today.

How could we detect these small and dense subhalos?  While a complete analysis of their observational signatures lies beyond the scope of this work, we briefly highlight a few promising detection avenues here. 
If the dark matter self-annihilates, these compact microhalos will be gamma-ray sources; they may be observed as point sources, and they would make a significant contribution to the observed gamma-ray background \cite{DMS05, OTN05, GGMDS07, SS09, SKM10, IME10, AMB10}.  Unfortunately, since the dark matter particles in these microhalos were created through scalar decay and cannot be thermal relics, there is no lower limit on their self-annihilation cross section.  
Numerous and dense subhalos near our location would also affect direct detection rates by changing the local dark matter density \cite{Kamion08, SKM10} and the direction of the dark matter wind \cite{AMB10}, which would alter both the event rate and its annual modulation.  If dark matter particles are directly detected, then the temporal correlation of the signal can be used to probe the abundance of local small subhalos because the event rate will change as a microhalo passes through the Solar System \cite{AMB10}. 

Of course, it is entirely possible that the dark matter particles in these microhalos do not self-annihilate and interact too weakly to be detected directly.  
The only guaranteed observational signatures of dark matter microhalos are gravitational.  Dark matter microhalos are gravitational lenses; we could hope to detect them through their impact on individual images of a strongly lensed quasar \cite{SW98, DMS05, CK10}.  Unfortunately, even though the early-forming microhalos that result from perturbation growth during reheating are more compact than their standard counterparts, their virial radii are still much larger than their Einstein radii, and it is unlikely that they could act as strong lenses.  Recently, it has suggested that we may be able to detect subhalos within our galaxy through their astrometric microlensing signatures \cite{EL11} or their impact on the pulse arrival times from millisecond pulsars \cite{BAZ11}.  These studies found that subhalos in the standard scenario are too rare and too diffuse to be detectable through these methods given our current level of astrometric and timing precision, but the abundant and dense microhalo population resulting from low-temperature reheating with direct dark matter production would significantly enhance both signals.  It is also possible that these microhalos may be dense enough to be detected through photometric microlensing \cite{RG09}.

Finally, we note that the Laser Interferometer Space Antenna (LISA) will be able to detect compact objects with masses greater than $10^{16}$ g that pass near its detectors; if these objects are the primary component of the dark matter, then a few such events are expected per decade \cite{SC04}.  LISA may also be sensitive to small and dense dark matter microhalos, if their radii are smaller than LISA's arm length ($\sim 5 \times 10^{11}$ cm) \cite{Berezinsky10}.  Microhalos that form before $z=340$ can have virial masses greater than $10^{16}$ g and virial radii less than $ 5 \times 10^{11}$ cm.  Such microhalos are less massive than the earth-mass microhalos that we have focused on and would form at high redshift for larger values of the reheat temperature.  For instance, if $T_\mathrm{RH} = 12$ GeV, then the critical mass at $z=500$ is $M_*=2.5\times10^{16}$ g, which is much less than the reheat horizon mass ($\mdec = 4\times 10^{19}$ g).  The Press-Schechter mass function predicts that 30\% of the dark matter is contained in halos with $M\simeq M_*$ if $M_*\ll\mdec$.  If these microhalos survive their absorption into larger halos, they could make up a significant fraction of the Galaxy's dark matter halo, and they could be detectable by LISA.  It would be interesting to explore this possibility, and the possibility of detecting early-forming microhalos through astrometric, photometric and timing microlensing in more detail; detecting these small structures or constraining their abundance would probe the origins of dark matter and the state of the Universe between inflation and nucleosynthesis.

\acknowledgments
 We thank Niayesh Afshordi, Robert Brandenberger,  Matthew Johnson, and especially Jonathan Braden for helpful discussions.
 We thank the Aspen Center for Physics for hospitality during the completion of parts of this work.
 This research was supported by the National Science and Engineering Research Council (NSERC) of Canada.  KS is supported in part by a NSERC of Canada Discovery Grant. 
 This research was supported by the Perimeter Institute for Theoretical Physics.  
 Research at the Perimeter Institute is supported by the Government of Canada through Industry Canada and by the Province of Ontario through the Ministry of Research and Innovation.

\appendix
\section{Derivation of perturbation equations}
\label{sec:perts_derivation}
We treat the oscillating scalar field, the radiation, and the dark matter as perfect fluids with energy momentum tensors
\beq
T^{\mu\nu}= (\rho+p)u^\mu u^\nu + p\, g^{\mu\nu},
\eeq
where $\rho$ and $p$ are the fluid's density and pressure, respectively, and $u^\mu \equiv \drm x^\mu/\drm\lambda$ is its four-velocity.  In this Appendix, we will take $x^0$ to be the proper time $t$.
The dark matter and the oscillating scalar fields are both pressureless fluids, while the radiation has $p = \rho/3$.  Since the scalar field is decaying into radiation and matter, these three fluids are exchanging energy, as described in Eqs.(\ref{bkgdS})-(\ref{bkgdM}).  We can express this energy exchange covariantly:
\beq
\nabla_\mu \left( ^{(i)}{T^{\mu}}_{\nu}\right) = Q^{(i)}_{\nu},
\label{cons}
\eeq
where $i$ denotes the individual fluids.  It follows from Eqs.(\ref{bkgdS})-(\ref{bkgdM}) that
\begin{subequations} 
\label{Qexp} 
\beqa
Q^{(\phi)}_{\nu} &=& ^{(\phi)}{T_{\mu\nu}}u_\phi^\mu\gam \\
Q^{(\mathrm{r})}_\nu &=& -(1-f) \, Q^{(\phi)}_{\nu}\\
Q^{(\mathrm{dm})}_\nu &=& -f \, Q^{(\phi)}_{\nu}.
\eeqa
\end{subequations}
Thus we see that, in our three-fluid model, 
\beq
Q^{(\phi)}_{\nu}+Q^{\mathrm{(r)}}_{\nu}+Q^{\mathrm{(dm)}}_{\nu} = 0,
\eeq
as required by the conservation of energy and momentum.

We obtain the perturbation equations by evaluating Eq.~(\ref{cons}) with the perturbed metric 
\beq
\drm s^2 = -(1+2\Psi)\drm t^2 + a^2(t)\delta_{ij}(1+2\Phi)\drm x^i \drm x^j
\eeq
and with perturbations in the density of each fluid: $\rho_{i}(t, \vec{x}) = \rho_i^0(t)[1+\delta_i(t, \vec{x})]$.  We also introduce perturbations to the four-velocity of each fluid: $u^0 = (1-\Psi)$ and $u^j_{(i)}= (1-\Psi)V_{(i)}^j$, where $V_{(i)}^j \equiv \drm x^j/\drm t$ is the fluid velocity of the $i$th fluid.  It follows that
\beqa
Q^{(\phi)}_{0} &=& \gam\rhos^0(1+\dels+\Psi)\\
Q^{(\phi)}_{j} &=& -\gam\rhos^0 a^2 \delta_{kj} V_\phi^k,
\eeqa
to first order in the perturbations.  Thus we see that $Q^{(\phi)}_{j}$ is a first-order quantity, while $Q^{(\phi)}_{0}$ has both a zero-order component $[Q^{(\phi),(0)}_{0}=\gam\rhos^0]$  and a first-order component $[Q^{(\phi),(1)}_{0}=\gam\rhos^0(\dels+\Psi)]$.

The $\mu=0$ component of Eq.~(\ref{cons}) implies that, for each fluid
\beq
\frac{\drm\delta}{\drm t}+ (1+w)\frac{\theta}{a} + 3 (1+w)\frac{\drm\Phi}{\drm t}= \frac{1}{\rho^0}\left[Q_0^{(0)} \delta -Q_0^{(1)}\right],
\eeq
where $w \equiv p/\rho$ is the fluid's equation of state parameter, $\theta \equiv a\, \partial_i V^i$ is the divergence of the fluid's conformal velocity, and $Q_0^{(0)}$ and $Q_0^{(1)}$ are the zero-order and first-order components of $Q_0$ for this fluid. The divergence of the spatial components of Eq.~(\ref{cons}) implies that
\beqa
\frac{\drm\theta}{\drm t} + (1-3w)H\theta + \frac{\nabla^2 \Psi}{a} + \frac{w}{1+w}\frac{\nabla^2 \delta}{a}=  \nonumber
\\
\frac{1}{\rho^0}\left[\frac{\partial_i Q_i}{a(1+w)} + Q_0^{(0)}\theta\right].\eeqa
Evaluating these equations for each fluid using Eq.~(\ref{Qexp}) yields the perturbation equations given by Eq.~(\ref{pertset}).

\section{Initial conditions for the Perturbations}
\label{sec:perts_ic}

The equation suite that we solve is
\begin{widetext}
\begin{subequations} 
\label{pertseta} 
\beqa
a^2 E(a) \dels^\prime(a) +\thest(a)+3a^2E(a)\Phi^\prime(a) &=& a \gamt \Phi(a), \label{delsa}\\
a^2 E(a) \thest^\prime(a)+aE(a)\thest+\tilk^2\Phi(a) &=&0, \label{thesa}\\
a^2 E(a) \delr^\prime(a)+\frac{4}{3}\thert(a)+4a^2 E(a) \Phi^\prime(a)&=& (1-f)\frac{\rhost^0(a)}{\rhort^0(a)}a\gamt\left[\dels(a)-\delr(a)-\Phi(a)\right], \label{delra}\\
a^2 E(a) \thert^\prime(a)+\tilk^2\Phi(a) -\tilk^2\frac{\delr(a)}{4}&=& (1-f)\frac{\rhost^0(a)}{\rhort^0(a)}a\gamt\left[\frac{3}{4}\thest(a)-\ther(a)\right], \label{thera}\\
a^2 E(a) \delm^\prime(a) +\themt(a)+3a^2E(a)\Phi^\prime(a) &=& f\frac{\rhost^0(a)}{\rhomt^0(a)}a\gamt\left[\dels(a)-\delm(a)-\Phi(a)\right], \label{delma}\\
a^2 E(a) \themt^\prime(a)+aE(a)\themt+\tilk^2\Phi(a) &=&  f\frac{\rhost^0(a)}{\rhomt^0(a)}a\gamt\left[\thest(a)-\themt(a)\right] \label{thema}\\
\tilk^2\Phi +3aE^2(a)\left[a^2\Phi^\prime(a)+a\Phi(a)\right] &=& \frac{3}{2}a^2\left[\rhost^0(a)\dels(a)+\rhort^0(a)\delr(a)+\rhomt^0\delm(a)\right], \label{phia}
\eeqa
\end{subequations}
\end{widetext}
where a prime denotes differentiation with respect to $a$.  

We chose the initial time $t_0$ such that all modes of interest are super-Hubble ($\tilk < 1$).  In conformal Newtonian gauge, super-Hubble perturbations do not evolve; to zeroth order in $\tilk^2$ we may set $\Phi^\prime = \dels^\prime = 0$.   By choosing $\gamt \ll1$, we have ensured that the Universe is initially dominated by the scalar field energy density, which implies that $E(a) = a^{-3/2}$ and $\rhost \gg \rhort, \rhomt$.  Also, $\rhost(t_0) \simeq 1$ implies that $\rhost \simeq a^{-3}$ prior to the decay of the scalar field.  With these conditions, and neglecting terms that are proportional to $\tilk^2$, and $\Phi^\prime$, Eq.~(\ref{phia}) implies that $\dels(t_0) = 2\Phi(t_0)$.  Equation~(\ref{thesa}) reduces to
\beq
\sqrt{a} \thest^\prime(a) +\frac{1}{\sqrt{a}}\thest +\tilk^2\Phi = 0.
\eeq
The solution to this equation is $\thest = -(2/3)\tilk^2\Phi\sqrt{a}$.

The solutions for $\rhom(t)$ and $\rhor(t)$ prior to the decay of the scalar field [Eqs.~(\ref{rhomsol}) and (\ref{rhorsol})] imply 
\beqa
(1-f)\frac{\rhost^0(a)}{\rhort^0(a)}a\gamt &=& \frac{5}{2\sqrt{a}}, \\
f\frac{\rhost^0(a)}{\rhomt^0(a)}a \gamt&=& \frac{3}{2\sqrt{a}}. \label{ratm}
\eeqa
Therefore, the right-hand sides of Eqs.~(\ref{delra}) and (\ref{delma}) are not negligible at $t=t_0$, even though $\gamt \ll1$.  The left-hand sides of Eqs.~(\ref{delra}) and (\ref{delma}) are initially zero, however, because the density perturbations and $\Phi$ do not evolve while the mode is superhorizon, and the velocity terms are proportional to $\tilk^2$.  To solve these equations at the initial time, we must set the right-hand sides of Eqs.~(\ref{delra}) and (\ref{delma}) to zero by demanding that $\delr=\delm = \dels - \Phi$.  Therefore, our first-order initial conditions for the matter and radiation perturbations are $\delr (t_0)=\delm(t_0) = \Phi(t_0)$. 

For superhorizon modes in a universe dominated by an oscillating scalar field, Eq.~(\ref{thera}) becomes
\beq
\sqrt{a} \thert^\prime(a) +2\tilk^2\Phi+\frac{5}{2\sqrt{a}}\thert = 0;
\eeq
we have used $\thest = -(2/3)\tilk^2\Phi\sqrt{a}$ and $\delr = \Phi$ to arrive at this equation.  The solution to this equation is $\thert = -(2/3)\tilk^2\Phi\sqrt{a}$.  For superhorizon modes in a universe dominated by an oscillating scalar field, Eq.~(\ref{thema}) also becomes
\beq
\sqrt{a} \themt^\prime(a) +2\tilk^2\Phi+\frac{5}{2\sqrt{a}}\themt = 0,
\eeq
where we have again used $\thest = -(2/3)\tilk^2\Phi\sqrt{a}$.  Since this is the same equation as we obtained for $\thert$, we have the same initial condition for $\themt$.

We can obtain more accurate initial conditions for the density perturbations by inserting the early-time solutions for the velocity perturbations into Eq.~(\ref{pertseta}).   Together, Eqs.~(\ref{phia}) and (\ref{delsa}) imply that the ${\cal O}(\tilk^2)$ term in the early-time solution for $\Phi(a)$ is proportional to $a^{-5/2}$ and has an undetermined coefficient.  Since this is a decaying mode that diverges as $a$ approaches zero, we choose this coefficient to be zero and continue to set $\Phi=\Phi_0$ and $\Phi^\prime(a)=0$ at early times.
If we do not neglect the $\tilk^2 \Phi$ term in Eq.~(\ref{phia}), we see that 
\beq
\dels(a) = 2\Phi_0 + \frac{2}{3}\tilk^2\Phi_0a
\eeq
at early times.  Using this solution and \mbox{$\thert=\themt= -(2/3)\tilk^2\Phi\sqrt{a}$} in Eqs.~(\ref{delra}) and (\ref{delma}) gives 
\beqa
\delr(a) &=& \Phi_0+\frac{46}{63}\tilk^2\Phi_0a \\
\delm(a) &=& \Phi_0+\frac{2}{3}\tilk^2\Phi_0a 
\eeqa
at early times. Like our initial condition for $\Phi$, these solutions also include undetermined decaying modes, and we set these modes to zero.  As long as the Universe is scalar-dominated, which implies $\gamt a \ll1$, these early-time solutions for the perturbations satisfy Eq.~(\ref{pertseta}), with one exception: there is a neglected ${\cal O}(\tilk^4 a)$ term in Eq.~(\ref{thera}).  Therefore, we expect that the radiation perturbation will deviate from this early-time solution when the mode enters the horizon.  The scalar and matter perturbations, however, will follow these solutions until the Universe becomes radiation dominated.

\section{Fitting Functions for the Transfer Function}
\label{sec:ABfit}
We want to find functions $A(k/\kdec)$ and $B(k/\kdec)$ such that
\beq
\delm(a, k) = \frac{10}{9} \Phi_0 \left[A \ln \left(\frac{B \,a}{a_\mathrm{hor}}\right)\right],
\label{mod3}
\eeq
where $k = a_\mathrm{hor}H(a_\mathrm{hor})$.  We know that $A=9.11$ and $B=0.594$ for $k/\kdec \lsim 1$.  We also know that $A$ and $B$ are given by Eq.~(\ref{mddelk}) for $k/\kdec \gsim 10$.  We found that the numerical solution for $\delm(a,k)$ for $0.05 \lsim k/\kdec \lsim 100$ is well-fit by the following expressions for $A(k/\kdec)$ and $B(k/\kdec)$:
\beqa
A(x) &=& \exp\left[\frac{0.609}{\{1+2.15(\ln x -1.52)^2\}^{1.38}}\right]\\
&&\times \left[9.11 \,{\cal S}(5.02-x) +\frac{3}{5}x^2 \, \,{\cal S}(x-5.02)\right]\nonumber\\
\ln B(x) &=& \ln(0.594)  \,{\cal S}(5.02-x) + \ln\left(\frac{e}{x^2}\right)  \,{\cal S}(x-5.02),\nonumber
\eeqa
where
\beq
{\cal S}(y) = \frac{1}{2}\left[ \tanh \left(\frac{y}{2}\right) +1\right]
\label{stepfunc}
\eeq
serves as a smooth step function.  We have tested these fitting functions for several values of $\kdec$ and $a$, and we find that Eq.~(\ref{mod3}) fits the numerical solution for $\delm(a, k)$ to within 5\% for $0.05 < k/\kdec$, provided that the modes have entered the Hubble horizon and $\mbox{$\delta \rhor \gg \delta \rhom$}$.

We also want to find a scale-dependent growth function $D(k, z)$ such that
\beq
\delm(k, z) = D_{50}(k,z) \times \delm(k, z=50).
\eeq
The function $D_{50}(k,z)$ should mimic the scale dependence of  $T_\mathrm{CAMB}/T_\mathrm{EH}$ at redshift $z=50$.  This ratio is approximately unity for $k/\keq \lsim 10^4$, and then it decreases to 0.773 for $k/\keq \gsim 10^5$.  This transition follows $T_\mathrm{CAMB}/T_\mathrm{EH} \simeq D_s(k)$, where
\beq
D_s(k) \equiv \frac{D_a-D_b}{1+\left(\frac{k/\keq}{48500}\right)^{2.1}}+D_b
\label{Dk}
\eeq
with $D_a = 1$ and $D_b=0.773$.  For redshifts \mbox{$3\lsim z \lsim 500$}, the modes with $T_\mathrm{CAMB}/T_\mathrm{EH} \simeq 1$ are proportional to $(2/3)+(a/a_\mathrm{eq})$, while the modes with \mbox{$T_\mathrm{CAMB}/T_\mathrm{EH} \simeq 0.773$} are proportional to ${\cal D}(a)$ as defined by Eq.~(\ref{fullD}).  Therefore, we define our scale-dependent growth function in this redshift range as \mbox{$D_{50}(k,z) = D_s(k)$} with
\beqa
D_a(z) &=& \frac{\frac{2}{3}+\frac{1+z_\mathrm{eq}}{1+z}}{\frac{2}{3}+\frac{1+z_\mathrm{eq}}{51}} \label{Da} \\
D_b(z) &=& \frac{{\cal D}(z)}{{\cal D}(z=50)}. \label{Db}
\eeqa
At lower redshifts, the dark energy slows the growth of density perturbations.  Since we expect the microhalos described in Section \ref{sec:structure} to be contained within larger structures at these redshifts, the transfer function at $z\lsim 3$ is not very important to our analysis.  To facilitate comparisons with the standard cosmological scenarios, however, we do report $T(k)$ and  $\sigma(M)$ evaluated at the present day, and so we need to extend our growth function to $z=0$.  We take
\beq
D_{50}(k,z) = \left\{ \begin{array}{cc}
D_s(k,z)& \mbox{ for } z \geq 2.67\\
\frac{D_1(z)}{D_1(z=2.67)}D_s(k,z=2.67) &\mbox{ for } z <2.67
\end{array}\right.
\eeq
where $D_s(k_z)$ is defined by Eqs.~(\ref{Dk})-(\ref{Db}), and 
\beq
D_1(z) = \frac{5 \Omega_\mathrm{M}}{2} \frac{H(z)}{H_0} \int_0^{(1+z)^{-1}} \frac{H_0^3}{[a H(a)]^3} \,\drm a.
\eeq
We chose $z=2.67$ to be the transition point because 
\beq
\frac{D_1(z=2.67)}{D_1(z=50)} = D_a(z=2.67).
\eeq
Consequently, on large scales ($k/\keq \lsim 10^4$), 
\beq
D_{50}(z<2.67) = \frac{D_1(z)}{D_1(z=50)}.
\eeq

\end{document}